\documentclass[hidelinks, 12pt]{article}

\usepackage[small,bf]{caption}
\usepackage{fullpage}
\usepackage[utf8]{inputenc}
\usepackage{appendix}
\usepackage{hyperref}
\usepackage{graphicx}
\usepackage{amsmath}
\usepackage{amssymb}
\usepackage{amsthm, amsfonts}
\usepackage{subfig}
\usepackage{float}
\usepackage{url}
\usepackage{tikz}

\let\phi\varphi
 
\newcommand{\da}{\Delta}
\newcommand{\db}{\Delta'}

\newcommand{\ra}{R}
\newcommand{\rb}{R'}

\newcommand{\downto}{\downarrow}
\newcommand{\upto}{\uparrow}

\newcommand{\ones}{\mathbf 1}
\newcommand{\reals}{{\mbox{\bf R}}}

\newcommand{\Prob}{\mathop{\bf Prob}}

\newcommand{\dom}{\mathop{\bf dom}} 

\newcommand{\eg}{{\it e.g.}}
\newcommand{\ie}{{\it i.e.}}

\newcommand{\BEAS}{\begin{eqnarray*}}
\newcommand{\EEAS}{\end{eqnarray*}}
\newcommand{\BEA}{\begin{eqnarray}}
\newcommand{\EEA}{\end{eqnarray}}
\newcommand{\BEQ}{\begin{equation}}
\newcommand{\EEQ}{\end{equation}}
\newcommand{\BIT}{\begin{itemize}}
\newcommand{\EIT}{\end{itemize}}

\title{When does the tail wag the dog? \\ Curvature and market making}
\author{Guillermo Angeris\\
{\small \texttt{angeris@stanford.edu}}
\and
Alex Evans\\
{\small \texttt{alex@placeholder.vc}}
\and
Tarun Chitra\\
{\small \texttt{tarun@gauntlet.network}}}

\date{December 2020}

\begin{document}

\maketitle

\begin{abstract}
Liquidity and trading activity on constant function market makers (CFMMs) such as Uniswap, Curve, and Balancer has grown significantly in the second half of 2020. 
Much of the growth of these protocols has been driven by incentivized pools or `yield farming', which reward participants in crypto assets for providing liquidity to CFMMs. 
As a result, CFMMs and associated protocols, which were historically very small markets, now constitute the most liquid trading venues for a large number of crypto assets.
But what does it mean for a CFMM to be the most liquid market?
In this paper, we propose a basic definition of price sensitivity and liquidity.
We show that this definition is tightly related to the curvature of a CFMM's trading function and can be used to explain a number of heuristic results. For example, we show that low-curvature markets are good for coins whose market value is approximately fixed and that high-curvature markets are better for liquidity providers when traders have an informational edge. Additionally, the results can also be used to model interacting markets and explain the rise of incentivized liquidity provision, also known as `yield farming.'

\end{abstract}

\section*{Introduction}
With the advent of Bitcoin and, more generally, the blockchain, there has been a strong desire for automated censorship-resistant decentralized exchanges (DEXs).
As blockchains provide a censorship-resistant means for executing programs in replicated state machines, many early DEX designs focused on emulating traditional market structure.
These early attempts implemented data structures from conventional markets, such as limit order books, in smart contracts.
However, due to both computational and latency constraints, blockchains often end up being suboptimal for order books. Additional notions for 

\paragraph{Constant function market makers.}
One solution to this problem is the family of constant function market makers (CFMM)~\cite{angeris2020improved}, starting with Uniswap~\cite{adams_2019, angeris2019analysis} which were invented as blockchain-native mechanisms for decentralized trading.
CFMMs require constant space and time interactions, unlike limit order books which have $O(n)$ space and, in many practical applications, $O(n^2)$ time complexity to process $n$ trades. (There are some implementations achieving asymptotically better results in space, such as~\cite{jenq2018order}, but these designs are not, to our knowledge, used in practice.)
This constant space and time requirement is ideal for blockchain environments where storage is expensive, as any data stored needs to be replicated and available at all consensus nodes, while compute can be costly for end users due to transaction fees. 

\paragraph{CFMM agents and interactions.} CFMMs are a special type of market maker that mediates the interactions between two principal agents: liquidity providers (LPs) and traders.
LPs provide capital to the CFMM by locking assets in a smart contract that implements the CFMM. LPs are then incentivized to not withdraw capital in order to earn trading fees from traders who trade against liquidity providers' locked capital.
This is implemented using the following mechanism: when a liquidity provider locks their assets into a CFMM smart contract, the smart contract creates LP shares and sends them back to the LP agent.
These LP shares are tokens that are essentially vouchers for the cash flows of the CFMM---they can be redeemed at a future time for the LP's share of the CFMM's assets and a pro-rata share of fees.
For instance, if an LP provides 10\% of the liquidity to a pool, then upon redemption they receive 10\% of the assets held in the pool, which includes the fees accrued by the pool.

\begin{figure}
    \centering
    \includegraphics[width=0.9\textwidth]{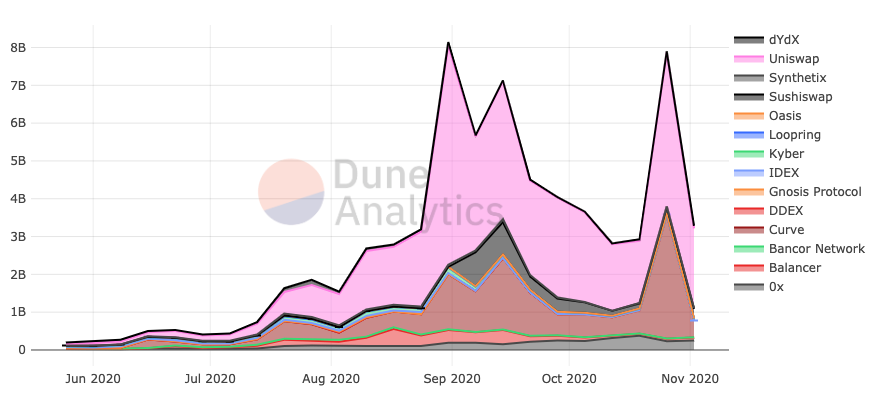}
    \caption{Decentralized exchange volumes in 2020. Uniswap has had dominant market share over the year, although it was challenged by Sushiswap in late-August. Source: Dune Analytics}
    \label{fig:dex_volumes}
\end{figure}

\paragraph{Relative liquidity of CFMMs.} Prior work on CFMMs~\cite{angeris2019analysis, angeris2020improved} has analyzed necessary and sufficient conditions for CFMMs to track an external reference market with infinite liquidity.
In this setting, CFMMs are modeled as secondary markets with finite liquidity whose price is adjusted by arbitrageurs to match that of the reference market.
This model allows one to answer the question of whether CFMMs can serve as price oracles; \ie, difficult-to-manipulate on-chain price feeds that other smart contracts can use.
However, the recent advent of incentivized CFMMs (referred to as ``yield farming") has led to a number of hundred million dollar markets whose \emph{reference} market is a CFMM with finite liquidity.
In fact, Uniswap's volume of \$440m on August 30, 2020 surpassed that of Coinbase Pro (\$380m), the largest US cryptocurrency market, making a number of markets significantly more liquid on CFMMs than on centralized order books~\cite{uni_coinbase_theblock, uni_coinbase_coindesk}. (See figure~\ref{fig:dex_volumes}.)
The natural next question to explore is: what happens when a CFMM becomes the \emph{most liquid} market; \ie, when the CFMM is more liquid than the external market?

In order to answer this question, we analyze how a CFMM market interacts with external markets 
that have finite liquidity. Our framework (\S\ref{sec:market_model} and~\S\ref{sec:model-description})
is general enough to include 
external markets that are limit order books or other CFMMs. For instance, the secondary market 
could be Uniswap and the external market could be a Balancer pool with the same assets. With this 
framework, we find that an analogue of the Gaussian curvature of a CFMM's trading function 
dictates how much prices differ between a CFMM market and a less-liquid secondary market, after 
no-arbitrage is enforced. Given that CFMM trading functions need only be convex and not smooth, we 
construct an analogue of standard Gaussian curvature that works in the convex setting.

\begin{figure}
    \centering
    \includegraphics[scale=0.45]{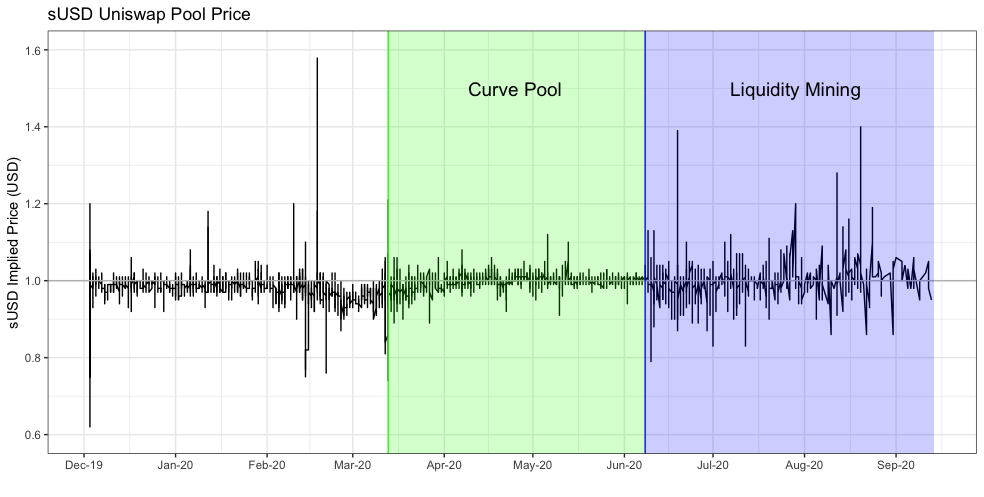}
    \caption{Implied price of sUSD on Uniswap. Shaded regions indicate the initial period after the Curve pool was deployed and the period of heavy incentivization of liquidity in on-chain CFMMs (``yield framing"), respectively. Source: Uniswap.}
    \label{fig:susd}
\end{figure}

\paragraph{Price stability.} In practice, highly liquid CFMMs appear to have local price stability 
between pairs of reserve assets. Issuers of on-chain assets have used this property to incentivize 
additional liquidity in low-curvature CFMMs in order to reduce volatility in the targeted assets. 
A well-known example involves sUSD, a dollar-pegged `stablecoin' issued through the Synthetix 
protocol~\cite{synthetix_mintr}. While sUSD is intended to track the price of \$1, in practice it 
was quite volatile around this peg, making it less useful to users seeking a stable coin price. In 
response, in March 2020, Synthetix incentivized the creation of a low-curvature CFMM through 
Curve~\cite{egorov_2019} that provided trading pairs between sUSD and other, more liquid, 
stablecoins. Shortly after the deployment of the CFMM, sUSD appeared significantly less volatile. 
Our framework provides a plausible explanation for this apparent price-stabilizing effect of
low-curvature CFMMs. Using the definitions provided in~\S\ref{sec:market_model} and a
basic inequality in~\S\ref{sec:price-stability}, we relate the price stability to the liquidity 
differences between the CFMM and the external market. While this model provides a plausible explanation for 
the stability in sUSD price observed between March and June 2020 (green shaded region in 
figure~\ref{fig:susd}), it does not explain the subsequent volatility experienced between June and 
September 2020 (purple region). This latter period corresponds to the rise of `yield farming' in 
the summer of 2020, wherein a number of stablecoins traded above their dollar pegs. We extend our 
model in~\S\ref{sec:yield_farming_returns} to incorporate the interaction between yield 
farming and curvature in CFMMs.

\paragraph{Liquidity provider returns.} A central question, answered 
in~\cite{angeris2020improved}, is to determine the expected value or payoff of holding LP shares 
under the assumption of no-arbitrage.
This payoff can be used to compare the returns that a CFMM liquidity provider earns to those of a market maker on 
a traditional exchange (\eg, a limit order book). Unlike limit order books, CFMMs only offer 
market orders and have deterministic slippage costs. The deterministic nature of slippage in CFMMs 
often leads to a variety of front-running attacks and deadweight loss that miners and validators 
can capture at the expense of users~\cite{daian2020flash}. However, this deterministic slippage 
also makes it easier for LPs to compute their expected payoff or expected loss.

Using some results from mathematical finance~\cite{carr2001towards}, it can be shown that 
Uniswap, the first CFMM to have over \$100 million in digital assets, has LP shares that should be 
thought of as closer to a perpetual options underwriter than a spot trading 
venue~\cite{clark2020replicating}. Combined, these properties suggest that 
CFMM LPs own a combination of derivative securities on the underlying tokens.
With CFMMs that can dynamically adjust, such as Balancer~\cite{balancer}, one can arbitrarily tune 
the payoff function of these derivative securities by adjusting the shape of the CFMM pricing 
curve~\cite{evans2020liquidity}. Given that these characteristics deviate from those of 
conventional spot financial exchanges, it is a natural to ask what happens when a CFMM is the 
dominant market for a set of assets.

\paragraph{Practical observations.} The intuition for why curvature affects actual market performance comes from empirical 
observations that CFMMs with lower curvature can increase LP profits for mean-reverting assets.
Curve~\cite{egorov_2019}, which was designed for trading highly correlated assets by offering 
lower curvature, was able to attract \$1 billion in liquidity and reach \$350 million in daily 
trading volume because LPs were able to extract more fees than in Uniswap. 
On the other hand, Balancer provides an adjustable CFMM curve, whose curvature controls the amount 
of loss engendered by LPs when large price moves occur.

\paragraph{Uninformed trading.} 
Using some basic results and definitions, we show in~\S\ref{sec:mean-revert} that
assets in which trades are mostly uninformed provide LPs with positive payoff whenever the curvature is low.
These results also suggest that the curvature of a CFMM trading function controls how much rebalancing a given CFMM will perform.
This provides a clear reason for why certain curves appear to be better in practice for mean-reverting assets (\eg, Curve for stablecoin--stablecoin trades).

\paragraph{Informed trading.} Next, we show that, in order for an LP to be profitable, the curvature of a CFMM should be large to account for the amount of information that an informed trader brings to the market.
This result is an analogue of classical market microstructure results involving multiplayer games between informed traders and market makers~\cite{kyle1985continuous, glosten1985bid, glosten1994electronic}.
Our formulation extends the informed trader framework for Uniswap of~\cite{aoyagi2020lazy} to general CFMMs through a two-player game where the informed trader makes a maximum bet~\cite{busseti2016risk} on the next market price update given an informational advantage.
Using the curvature framework of \S\ref{sec:market_model}, we illustrate a condition (similar to Glosten's classical bound~\cite{glosten1994electronic}) that connects the informational edge of the informed trader and the curvature and fees of a CFMM to the payoffs of the informed trader and the LP. We extend this single period model to a multiperiod model and make conjectures about the optimal information flow in a CFMM in appendix~\ref{app:multiperiod_it}.

\paragraph{Yield farming.} We also show lower bounds to so-called `yield farming' payoffs that are sufficient to compensate LPs, based on the curvature difference between two markets. Yield farming can be seen as analogous to market-maker subsidies in conventional markets~\cite{brolley2013informed}. In yield farming, an LP first supplies reserves to a CFMM containing some token $T$ and a num\'eraire such as Ethereum (ETH). The LP then locks the corresponding LP shares they receive from the CFMM and receives newly-minted tokens of $T$ over time in return. By locking LP shares in the smart contract, users that provably provide liquidity for the T--ETH trading pair are subsidized.
In practice, this incentive bootstraps liquidity for token T by incentivizing users to be LPs~\cite{yield_farming_coindesk}.
We show that these incentives can be bounded from below by the curvature of a CFMM.

\paragraph{Portfolio Greeks.} Finally, we apply our results regarding curvature to analyze financial properties of CFMM LP shares.
First, we extend the frameworks of~\cite{angeris2020improved, clark2020replicating} to compute the first and second order Greeks (\eg, $\Delta, \Gamma$) for CFMM LP share payoffs in the two coin case, and then use the basic liquidity framework provided in~\S\ref{sec:market_model} to
give bounds on the portfolio Greeks. While we note that these results, due to the assumptions, may be of limited practical importance, they can be used to interpret the curvature bounds in several different ways. We leave a suitable and practical version of the result as an open conjecture.

\paragraph{Summary.} These results show that curvature is a crucial design parameter to tune when designing CFMMs.
A number of CFMM designs have been proposed for prediction markets~\cite{gnosis_amm}, derivatives trading~\cite{perpetual_protocol, mcdex}, and self-balancing ETFs~\cite{balancer}.
In each of these applications, a CFMM LP share represents a complex payoff function that changes dramatically based on the expected trading behavior of the assets held within the CFMM.
Our results show that the returns and payoffs realized by holders of CFMM LP shares are intrinsically tied to the curvature of the trading function.
Moreover, they explain a number of empirical outcomes that have happened throughout the large number of different CFMMs available on Ethereum.
CFMM designers need to be cognizant of trade-offs that are made by adjusting curvature, especially as payoff functions become increasingly complex.
These results also provide guidance on how to parameterize yield farming incentives to achieve certain liquidity targets.
As a number of yield farming assets have failed due to over-incentivization of liquidity~\cite{sushiswap_migration, yieldfarming_dead}, it is increasingly important to understand how to efficiently incentivize on-chain liquidity. 
These results provide ways to sensibly optimize incentives to meet liquidity goals.

\section{Two asset market model}\label{sec:market_model}
In this section, we define basic terms and models used throughout the remainder of the paper. In particular,
we will define what it means for a market to be `stable' or `liquid.' We note that, while some markets in practice can
simultaneously trade $n$ coins for $n$ coins, we will focus on the case where the market only trades two coins, and
this is the model we will use in the majority of the paper. (We will sometimes present the $n$ coin generalizations, as in, \eg,
appendices~\ref{app:ncoinpv} and~\ref{app:superreplication}, when they are simple, but this is the exception rather than the rule.) In 
this case, we will call one asset the \emph{traded coin}, and this asset is distinct from the \emph{num\'eraire}, which is the base 
currency used to measure prices. Unless otherwise specified, all trades (which we will often simply call $\Delta$) are positive when they are 
buying $\Delta$ amount of the traded asset, and negative when they are selling it.

\paragraph{Price impact function.} We will define the \emph{price impact} function $g: \reals \to \reals_{++}$ of a market to be the function that connects the market's marginal price before the trade, which we will call $m_0$, to the price after the trade. More specifically, we have that $g(0) = m_0$, while $g(-\Delta)$ specifies the CFMM's marginal price immediately after the trade, which sells $\Delta$ of the traded coin to the market, is performed.

We assume two basic facts about $g$: one, that $g$ is continuous, and, two, that $g$ is nondecreasing. In other words, $g$ is a continuous function that expresses how the market's price changes after a given trade, with the assumption that trades of size 0 (\ie, null trades) don't change the market price, and that larger trades lead to higher marginal prices. Note that these assumptions are common in the order-book literature (see, \eg,~\cite{bouchaud2004fluctuations}) and true for all CFMMs (see, \eg,~\cite{angeris2020improved}). Additionally, this assumption is equivalent to the convexity of the quantity function $q(\Delta) = \int_0^\Delta g(t)\,dt$, which is also a common assumption in the classical economics literature (see~\cite{gatheral2010no}).

\paragraph{No-arbitrage.} A common way to model interactions between different markets is the assumption of no-arbitrage. One
way of stating this assumption is through the existence of an agent, called an \emph{arbitrageur}. This agent is allowed to borrow
any amount of coin $\Delta$, trade it between the available markets (to receive some amount of coin, say, $\Delta^a$), and then pay back the borrowed amount $\Delta$, to receive $\Delta^a - \Delta$ profit. If there exists a trade which guarantees that $\Delta^a > \Delta$, we then say that there is an \emph{arbitrage opportunity} which we assume the arbitrageur will execute to receive strictly positive profit.

In our presentation, we will assume that there is an (infinitely-liquid) reference market with fixed price $m_a$. (We provide a generalization to reference markets which do not have infinite liquidity in~\S\ref{sec:price-stability}.) An arbitrageur would then attempt to maximize its profits by trading some amount of coin $\Delta$ between both markets, which would give a total payoff of
\[
\int_{0}^\Delta (m_a - g(-t))\,dt.
\]
A necessary condition for this quantity to be maximized is that the marginal price of both markets after the no-arbitrage trade, which we will call $\Delta^\star$, must be equal; \ie, that $\Delta^\star$ must satisfy
\begin{equation}\label{eq:noarb_condition}
m_a = g(-\Delta^\star),
\end{equation}
which follows from the first-order optimality conditions applied to the arbitrageur's payoff.

Without loss of generality, we will assume that $m_a \le m_0 = g(0)$. Note that this is always possible in a two-coin
economy by changing the choice of the coins to be traded; \ie, by swapping the places of the traded coin and the num\'eraire, the resulting market prices are $1/m_a \le 1/m_0$, which satisfy the above inequality. Because of this assumption and the fact that $g$ is a nondecreasing function, we will have that $\Delta^\star \ge 0$. (We discuss this further in the case where the market is not infinitely liquid in~\S\ref{sec:model-description}.)

\paragraph{Price stability.} We will say that the price impact function $g$ is $\mu$-\emph{stable} (with $\mu \ge 0$) if it satisfies:
\begin{equation}\label{eq:mu-stability}
g(0) - g(-\Delta) \le \mu \Delta.
\end{equation}
In other words, we say that the price impact function $g$ for some market is $\mu$-stable whenever a (nonnegative) trade of size $\Delta$ does not change the market's price by more than $\mu\Delta$. Because $g$ is increasing by assumption, when $\Delta \ge 0$, inequality~\eqref{eq:mu-stability} is equivalent to the sensitivity-like bounds,
\[
|g(0) - g(-\Delta)| \le \mu \Delta,
\]
as both sides are nonnegative.

There are several useful sufficient conditions for~\eqref{eq:mu-stability} to hold. For example, it suffices that its first derivative is bounded from above by $\mu$ for all $\Delta \ge 0$; \ie,
\[
\frac{\partial g(-\Delta)}{\partial \Delta} \le \mu,
\]
but this is not a necessary condition, as the function $g$ need not be differentiable in its second argument. We show a 
connection between this sufficient condition and the curvature of a trading function for a given CFMM in appendix~\ref{app:curvcfmm}; we also discuss explicit bounds for $\mu$ for some CFMMs used in practice, and how they relate to common intuition, in the following section.

\paragraph{Liquidity bounds.} We will say that a price impact function for a market is $\kappa$-\emph{liquid} if it satisfies
\begin{equation}\label{eq:k-liquidity}
g(0) - g(-\Delta) \ge \kappa \Delta.
\end{equation}
In other words, selling $\Delta$ coin to the market decreases the reported price by at least $\kappa\Delta$, which implies that there is some amount of price slippage that is linearly bounded from below by a factor of at least $\kappa$. Additionally, we note that it is possible (and, in fact, common for many markets) that a trading function is both $\mu$-stable and $\kappa$-liquid, with $\kappa \le \mu$.

\paragraph{Trade sizes.} It is often the case that bounds of the form of~\eqref{eq:k-liquidity} (and, sometimes, bounds
of the form of~\eqref{eq:mu-stability}, as we will see later in this section) are not global, but, instead, hold over some interval of size $L$; \ie, such bounds only hold for $0 \le \Delta \le L$. In these specific cases, we will mention the corresponding conditions in the statements, as required for the results to hold for CFMMs in practice. On the other hand, we note that all of the proofs presented have results which immediately carry over in this case, even when the interval is not explicitly mentioned. (We leave such extensions as simple exercises for the reader.)

\subsection{CFMM curvature}\label{sec:cfmm-curvature}
In decentralized finance, the market we are studying is almost always a \emph{constant function market maker} or CFMM. (See, \eg,~\cite{angeris2020improved} for an introduction.)
In this case, the markets' behaviors are specified by (often simple) mathematical formulae, and often have closed-form solutions for the constants $\mu$ and $\kappa$. In this section, we will show how to compute $\mu$ and $\kappa$ for some of the CFMMs used in practice.

\paragraph{Constant function market makers.} A CFMM is an algorithmic market maker~\cite{othman2010automated, othman2013practical, hanson2003combinatorial} defined by its \emph{reserves}, specifying how much of each coin is available
for trading, and its \emph{trading function}, which controls whether the market maker will accept or reject a proposed trade. The reserves are given by $R \in \reals_+$ for the coin to be traded, and $R' \in \reals_+$ for the num\'eraire coin,
while its trading function is given by $\psi: \reals_+^2\times \reals^2 \to \reals$. The trading function maps the pair of reserves $(R, R') \in \reals_+^2$ and a trade, purchasing $\Delta$ of the coin to be traded and $\Delta'$ of the num\'eraire, $(\Delta, \Delta') \in \reals^2$, to a scalar value.

When a CFMM has reserves $(R, R')$, any agent may propose a trade $(\Delta, \Delta')$. By definition, the CFMM accepts the trade
whenever
\[
\psi(R, R', \Delta, \Delta') = \psi(R, R', 0, 0).
\]
The CFMM then pays out $\Delta$ of the traded coin to the trader and receives $\Delta'$ of the num\'eraire. This results in the following  
update to the reserves: $R \gets R - \Delta$ and $R' \gets R' + \Delta'$. (Negative values of $\Delta$ and $\Delta'$ reverse the flow of 
the coin.)

For notational convenience, we will abuse notation slightly by writing $\psi(\Delta, \Delta')$ for $\psi(R, R', \Delta, \Delta')$,
such that the function $\psi(\cdot, \cdot)$ implicitly depends on the reserve values $(R, R')$ in the remainder of the paper. Additionally, because
our focus is on the two coin case and not the general $n$ coin case, we use different notation than~\cite{angeris2020improved} to
prevent overly-cumbersome proofs and results. We show the exact connection between both forms in appendix~\ref{app:notation}.

\paragraph{Marginal prices.} Given a CFMM with trading function $\psi$, the marginal price at these reserves is given by~\cite[\S2.4]{angeris2020improved}:
\begin{equation}\label{eq:g_def}
g(\Delta) = -\frac{\partial_1\psi(\Delta, \Delta')}{\partial_2\psi(\Delta, \Delta')},
\end{equation}
where $\partial_i\psi$ denotes the partial derivative of $\psi$ with respect to the $i$th argument and $\Delta'\in \reals$ is the (usually unique) solution to
\[
\psi(\Delta, \Delta') = \psi(0, 0),
\]
for a given $\Delta \in \reals$. Note that this is only defined whenever $\Delta \le R$ and $\Delta' \ge -R'$; \ie, when there are enough reserves to complete the trade. Because it is often the case that $\psi$ satisfies the condition $\Delta' = \int_0^\Delta g(-t)\,dt \le R'$, even as $\Delta \upto\infty$, we will consider these bounds to be implicit, unless otherwise stated.

\paragraph{Marginal prices with fees.} While it is possible to implicitly include fees in the definition of the CFMM's trading function,
it is often simpler to include the fee explicitly. In many cases such as Uniswap, Balancer, and Curve, the fee is given by some number
$0 < \gamma \le 1$ such that $(1-\gamma)$ is the percentage fee taken for each trade, and the fee-less CFMM, with trading function
$\psi$, is modified in the following way~\cite[\S3.1]{angeris2020improved}:
\[
\psi^f(\Delta, \Delta') = \psi(\gamma\Delta, \Delta') = \psi(0, 0),
\]
where $\psi^f$ is the CFMM trading function with fees, for trades which sell some amount of coin $\Delta \le 0$ to the CFMM. The reserves are updated in a similar way as the original CFMM. We note that the case where $\Delta \ge 0$ can be derived by appropriately exchanging the traded coin and the num\'eraire. The directionality here comes from the fact that fees are usually charged `on the way in,' or, in other words, asymmetrically charged to the coin being sold to the CFMM. (See, \eg, appendix~\ref{app:notation}.) 

In this case, we can write the marginal price of a given trade of size $\Delta \le 0$ after fees in terms of the marginal price
of the original CFMM, since
\[
g^f(\Delta) = -\frac{\partial_1\psi^f(\Delta, \Delta')}{\partial_2\psi^f(\Delta, \Delta')} = -\frac{\partial_1\psi(\gamma\Delta, \Delta')}{\partial_2\psi(\gamma\Delta, \Delta')} = \gamma g(\gamma \Delta),
\]
where $\Delta'$ is the (usually unique) solution to $\psi(\gamma\Delta, \Delta') = \psi(0, 0)$, and, as before, $\Delta \le 0$; \ie, we are selling the coin to be traded to the CFMM.

Given that we can express the price impact function $g^f$ with fees in term of the fee-less price impact function $g$, the next problem is to find bounds of the form of~\eqref{eq:mu-stability} for fee-less CFMMs, which we show for a few special cases.

\paragraph{Constant sum market maker.} The simplest example of a CFMM is the constant sum market maker, whose trading function
is the \emph{constant sum} trading function:
\[
\psi(\Delta, \Delta') = (R-\Delta) + (R'+\Delta').
\]
In this case, the marginal price is, from~\eqref{eq:g_def}:
\[
g(\Delta) = 1,
\]
whenever $-R' \le \Delta \le R$ and is otherwise undefined. In this case, we have the following bound, whenever
$0 \le \Delta \le R'$:
\[
g(0) - g(-\Delta) = 0 \le \mu \Delta,
\]
with curvature bound $\mu = 0$. (This is simply due to the fact that a constant sum market maker always reports a fixed price,
so long as it has nonzero reserves.) Similarly, we also have $\kappa = 0$, on the same interval, by the same argument.


\paragraph{Global bounds on $\mu$ for convex impact.} While the constant sum market maker has a simple enough
trading function that it can be analyzed directly, analyzing other trading functions in the same way can quickly lead to very
complicated results. A useful and simple condition applies in the common case that the price impact
function, $g$, is a differentiable convex function. In this case, we have that, for all valid $\Delta$,
\[
g(-\Delta) \ge g(0) - g'(0)\Delta,
\]
where $g'(0)$ is the derivative of $g$ evaluated at zero, by the first-order condition for convexity~\cite[\S3.1.3]{cvxbook}. This can be written as
\begin{equation}\label{eq:convex-bounds}
g(0) - g(-\Delta) \le g'(0)\Delta.
\end{equation}
Setting $\mu = g'(0)$ yields the desired result. When $g$ is differentiable, taking the limit as $\Delta \downto 0$, shows that this is
the tightest possible bound on $\mu$ over any nonzero interval size. (If $g$ is not differentiable, we can take $\mu$ to be the largest 
subgradient of $g$ at $0$, which is the tightest possible bound by the same argument.)

\paragraph{Local bounds on $\kappa$ for convex impact.} On the other hand, given any interval of size $L \ge 0$, we can give a $\kappa$-liquidity bound for $g$, by noting that, because
$g$ is convex, the definition of convexity gives the following inequality:
\[
g(-\Delta) = g\left(\left(1-\frac{\Delta}{L}\right)0 + \frac{\Delta}{L}(-L)\right) \le \left(1-\frac{\Delta}{L}\right)g(0) + \frac{\Delta}{L}g(-L).
\]
A basic rearrangement shows:
\[
g(0) - g(-\Delta) \ge \left(\frac{g(0) - g(-L)}{L}\right)\Delta,
\]
and setting $\kappa = (g(0) - g(-L))/L$ gives the result. Since the bound is tight at $\Delta = 0$ and $\Delta = L$,
this $\kappa$ yields the tightest possible bound along this interval.



\paragraph{Uniswap.} One of the simplest nontrivial convex bounds is the bound for Uniswap (or constant product market maker) with no fees, where we can write
\begin{equation}\label{eq:uniswap-marginal}
g(\Delta) = \frac{k}{(R - \Delta)^2},
\end{equation}
where $k = RR'$ is the product constant~\cite[App.\ A]{angeris2019analysis}.

If $R > \Delta$ (\ie, there are enough reserves to carry out a trade of size $\Delta$) this function is a convex function, as $x \mapsto 1/x^2$ is convex over the positive reals.
Using inequality~\eqref{eq:convex-bounds}, we have, for all $\Delta \ge 0$,
\[
g(0) - g(-\Delta) \le 2\frac{k}{R^3}\Delta = 2\frac{g(0)}{R}\Delta = \mu\Delta.
\]
In the special case where the marginal price at the zero trade is $g(0) = 1$ (as is common with stablecoin--stablecoin markets), which happens when $R = R'$, we can re-write $\mu$ in terms of the portfolio value of the reserves as
\[
\mu = \frac{4}{P_V},
\]
where the portfolio value is given by $P_V=g(0)R + R' = 2R$.

Similarly, for any interval size $L \ge 0$, we have
\[
\kappa = \frac{k}{L}\left(\frac{1}{R^2} - \frac{1}{(R+L)^2}\right) = \frac{g(0)}{L}\left(1 -\frac{1}{(1 + L/R)^2}\right).
\]
Note that both $\mu$ and $\kappa$ both decrease when $R$ increases as liquidity is added (\eg, via Uniswap's \verb|addLiquidity| function) for a fixed price $g(0)$ and interval size $L$. In other words, Uniswap's effective curvature \emph{decreases} as the reserves increase, as one might intuitively expect.
We can interpret this as, for a fixed trade cost, larger trades can take place in Uniswap with higher reserves. An alternative interpretation is that the cost of manipulation also increases in the reserve size $R$.
These results were proven in~\cite[\S2.3]{angeris2019analysis} using different techniques which do not easily generalize.

\paragraph{Two asset Balancer.} For Balancer (which is also sometimes called the constant mean market maker~\cite[\S3]{angeris2019analysis}, or the geometric mean market maker~\cite{evans2020liquidity}) with two assets and weight $\tau \in (0, 1)$, we have the trading function
\[
\psi(\Delta, \Delta') = (R - \Delta)^{\tau}(R' + \Delta')^{1-\tau}.
\]
Let $\xi = \frac{\tau}{1-\tau}$ for notational convenience. We then have:
\[
g(\Delta) = \frac{d \Delta'}{d\Delta} = \left(\frac{\tau}{1-\tau}\right) \frac{k^{1/(1-\tau)}}{(R-\Delta)^{1 + \tau/(1-\tau)}} = \frac{\xi k^{\xi/\tau}}{(R-\Delta)^{1+\xi}}.
\]
Note that when $\tau=\frac{1}{2}$ then $\xi = 1$ and this price impact function is equal to Uniswap's,
generalizing result~\eqref{eq:uniswap-marginal}.
This function is convex since $x \mapsto x^{-(1+\xi)}$ is convex over the positive reals for any $\xi > -1$. This implies
\[
g(0)-g(-\Delta) \leq \xi(1+\xi)\frac{ k^{\xi/\tau}}{R^{2+\xi}}\Delta = (1+\xi)\frac{g(0)}{R}\Delta = \mu\Delta.
\]
As with Uniswap, we can give a simple expression for $\mu$ in terms of the portfolio value in the special case where $g(0) = 1$:
\[
\mu=\frac{1}{\tau (1-\tau) P_V}
\]
where the portfolio value is given by $P_V = g(0)R + R' = R/\tau$. (Note that the expression for $\mu$ is symmetric about $\tau=1/2$,
even though the portfolio value $P_V$ is not.)

Similarly, for any interval of size $L$, we have that
\[
\kappa = \frac{\xi k^{\xi/\tau}}{L}\left(\frac{1}{R^{1+\xi}} - \frac{1}{(R+L)^{1+\xi}}\right) = \frac{g(0)}{L}\left(1 - \frac{1}{(1 + L/R)^{1+\xi}}\right)
\]
As before, we have that both $\mu$ and $\kappa$ are decreasing functions in the reserves $R$ for a fixed marginal price $g(0)$ and interval length $L$. We can additionally recover the Uniswap bounds on $\kappa$ and $\mu$ by setting $\tau = 1/2$, or, equivalently, $\xi = 1$.

\begin{figure}
    \centering
    \includegraphics[scale=0.5]{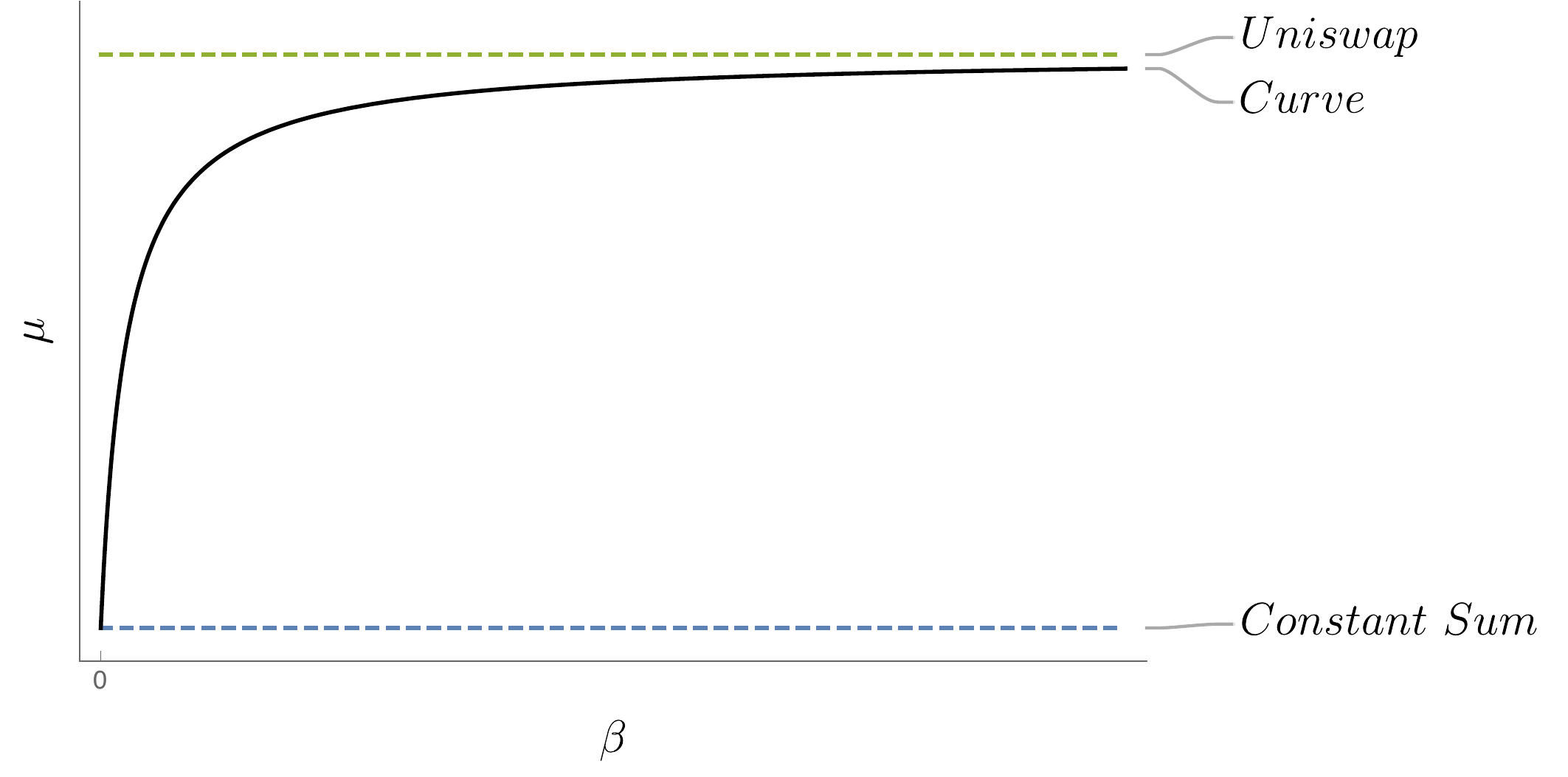}
    \caption{Curvature constant $\mu$ for Curve for different values of $\beta$, where $g(0) = 1$.}
    \label{fig:curve-mu}
\end{figure}

\paragraph{Two asset Curve.} Another very popular CFMM is Curve~\cite{egorovStableSwapEfficientMechanism}, with trading function
\[
\psi(\Delta, \Delta') = \alpha ((R - \Delta) + (R'+\Delta')) - \beta ((R-\Delta)(R'+\Delta'))^{-1}.
\]
It is worth noting that, as $\alpha/\beta$ becomes large, then $\psi$ is approximately close to the trading function
for the constant sum market maker. Similarly, as $\alpha/\beta$ becomes small, $\psi$ becomes similar to the Uniswap trading
function. (See, \eg, figure~\ref{fig:curve-mu}, which shows that $\mu$ for Curve converges to the curvature constants for constant sum
as $\beta \downto 0$ and Uniswap as $\beta \upto \infty$.) One can imagine variations on Curve where the product term is replaced by the
reciprocal of the weighted geometric
mean, as with Balancer, or a number of other functions.

\begin{figure}
    \centering
    \includegraphics[scale=0.5]{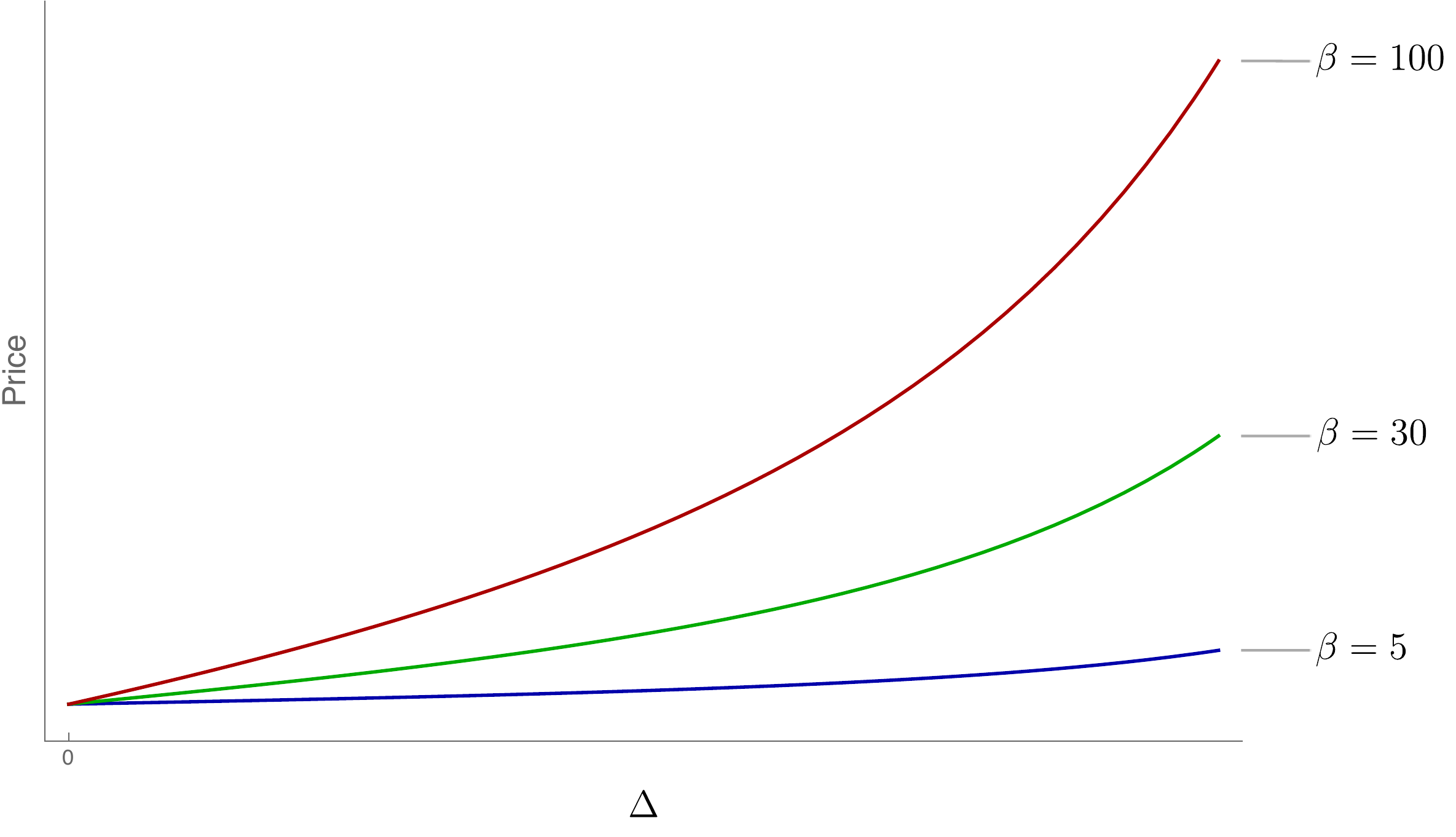}
    \caption{Curve price impact function plotted for different values of $\beta$. Here, $g(0) = 1$.}
    \label{fig:curve-convex}
\end{figure}

The marginal price function for Curve is relatively complicated, but can be derived with some work (see~\cite[\S2.4]{angeris2020improved}):
\[
g(\Delta) = \frac{\alpha(R - \Delta)(R' + \Delta') + \beta(R - \Delta)}{\alpha(R - \Delta)(R' + \Delta') + \beta (R' + \Delta')},
\]
where
\[
\Delta' = R' + \frac{1}{2\alpha}\left(\sqrt{(\alpha(R -\Delta) - k)^2 - 4\alpha\beta (R - \Delta)^{-1}} - (\alpha(R - \Delta) -k) \right),
\]
and we have defined $k= \psi(0, 0) = \alpha(R + R') - \beta (RR')^{-1}$ for notational convenience. From figure~\ref{fig:curve-convex} we see
that $g$ is indeed convex for a number of parameters $\beta$ (setting $\alpha = 1$ without loss of generality, as $g$ is homogeneous of degree zero with respect to $(\alpha, \beta)$). Showing that $g$ is convex is rather more involved; we provide a proof in appendix~\ref{app:curve_convex}.

While, in general, $\mu=g'(0)$ can yield complicated expressions for Curve, the special case where $g(0)=1$ can be written as a simple function of the portfolio value:
\[
\mu = \frac{32\beta}{8\beta P_V+ \alpha P_V^4},
\]
where $P_V = g(0)R + R' = 2R$. This provides a simple expression for the maximum slippage a trader can expect for a given trade size when assets on Curve are trading at their peg.

\begin{figure}
    \centering
    \includegraphics[scale=0.375]{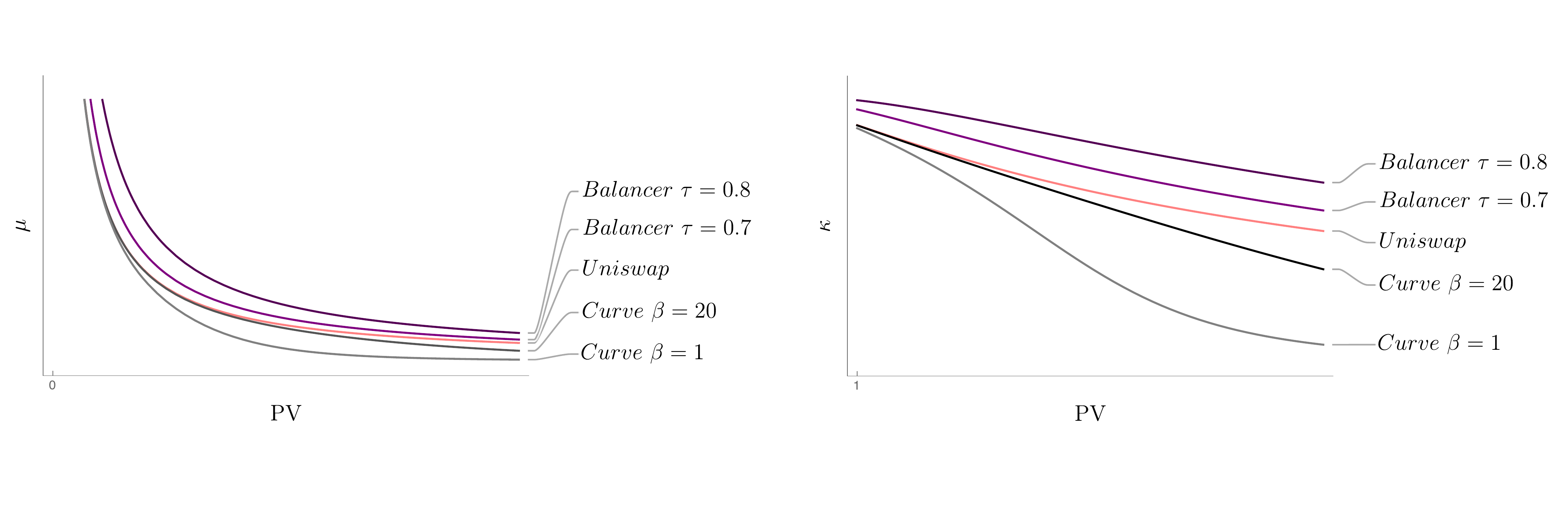}
    \caption{(Left) Plots of $\mu$ for Uniswap, Balancer, and Curve for different parameter choices as a function of the total reserve value for $g(0)=1$. (Right) Plots of $\kappa$ for Uniswap, Balancer, and Curve as a function of the total reserve quantity for $g(0)=1$ fixed and $L=1$.}
    \label{fig:normalized-curvature}
\end{figure}


\subsection{Examples in practice}
\begin{figure}
    \centering
    \includegraphics[scale=0.35]{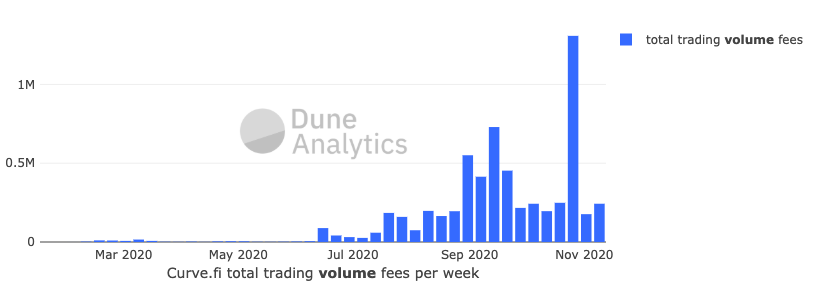}
    \includegraphics[scale=0.35]{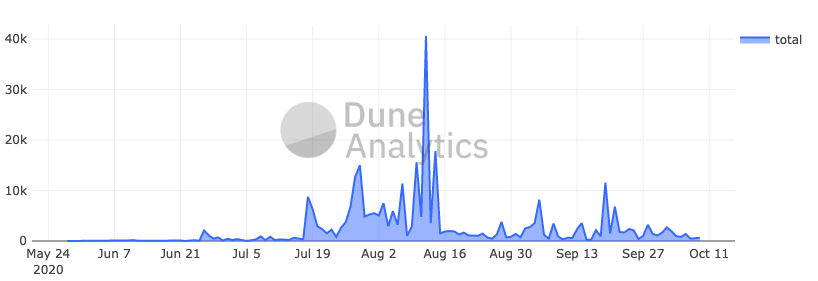}
    \caption{Aggregated trading fees for Curve (top) and mStable (bottom) in USD. Source: Dune Analytics.}
    \label{fig:curve-mstable-comparison}
\end{figure}

One of the simplest examples of curvature impacting trading performance is in stablecoin trading venues.
Stablecoins, which are assets that aim to be approximately pegged to a fiat num\'eraire such as the US dollar, are assets with an approximately constant price due to their peg.
However, their prices and outstanding supply often differ for systematic reasons.
For instance, one type of asset might only be centralized and only allowed to be created by non-US entities (\eg, Tether), whereas another asset is more decentralized (\eg, MakerDAO).
The former stablecoin might have a small number of large institutions using the coin whereas the latter is likely to have more small participants.
If this is the case, it will naturally be easier to perform bigger trades in the former currency. A CFMM designed for stablecoin--stablecoin trading, such as Curve, needs to have curvature that is adapted for these types of trading.

\paragraph{Uniformed trading.} Generally speaking, the trading of stablecoins for stablecoins tends to be uninformed.
That is, users often trade these stablecoin pairs because a specific smart contract or exchange that they want to interact with only allows the use of a particular stablecoin.
There is no information about the direction of trading and, given the ease of creation-redemption arbitrage for stablecoins, there is little use in trying to predict stablecoin prices~\cite{stablecoins_stable, oraclepricing}.

\paragraph{Curvature.} Intuitively, then, venues for stablecoin--stablecoin trades should have relatively low-curvature price impact functions, which would entice traders due to the small price slippage, and entice liquidity providers due to the small opportunity cost.
Taking this idea to its extreme, one might argue that assets that are supposed to be the same value should be traded on a CFMM with zero curvature.
An example of such a market is mStable, which uses the constant sum trading function presented in~\S\ref{sec:cfmm-curvature}.
These curvature-less markets have trouble responding to price, as they effectively quote a fixed price for any trade performed.
In practice, as illustrated in figure~\ref{fig:curve-mstable-comparison}, we see that Curve generates almost an order of magnitude more in trading fees for LPs than mStable. 
This is driven by two phenomena. First, a zero-curvature AMM will end up less liquid in practice because it quickly runs out of reserves as price fluctuates. In the case of mStable, there is a chronic shortage of Dai for trades as Dai is frequently trading above its peg.
Second, LPs face maximal opportunity cost relative to a low-curvature CFMM with the same assets and fees.
Most of the trading volume and liquidity provision on mStable appears instead to be driven by \emph{yield farming}. We discuss such incentives in more detail later in~\S\ref{sec:yield_farming_returns}.

\section{LP returns and curvature}
Empirically, it has been observed that returns to LPs in CFMMs are closely tied to both the shape of the CFMM trading set and the properties of the price process of the two assets.
Curve's advantage over Uniswap for mean-reverting, low volatility assets led to it attracting significantly more trading volume for certain assets.
As shown in \S\ref{sec:cfmm-curvature}, Curve has a low curvature regime around a particular price and a high curvature region far away from this price.
This design was chosen to optimize profits earned by LPs for mean reverting assets while allowing traders to place large sized orders when assets are near their mean.
A natural question to ask is: how much does adjusting the curvature of a CFMM for such assets affect LP returns?

\paragraph{Liquidity provider portfolio value.} The LP portfolio value is defined as the value of coins that an LP has locked in a CFMM.
If a user owns $b \in [0,1]$ percent of the LP shares in a CFMM, then they own the right to claim $bR$ of asset 1 and $bR'$ of asset 2, where $R$ and $R'$ are, as before, the reserves of the CFMM.
In this subsection, we will formalize heuristics arguments of~\cite{egorovStableSwapEfficientMechanism, egorov_2020} which show that LP portfolio values are directly affected by the curvature of a trading function.
We will also generalize these claims to generic price processes interacting with CFMMs by considering adverse selection towards LPs. In particular, we will consider how LP returns are affected by \emph{informed traders}, who have an estimate for the probability distribution of future prices. These results will illustrate that the design of efficient CFMMs for a variety of markets depends on how the curvature is adjusted to ensure that LPs can be profitable. In particular, we will see that, for stablecoins, where most  trades are uninformed, low curvature improves performance and LPs still come out with positive profit even for large trades, while markets where there exist informed traders with a bigger edge, higher curvature is preferable to prevent downside LP losses.
    
\subsection{Uninformed trading and low curvature CFMMs}\label{sec:mean-revert}
In this scenario, we consider a trader who wishes to buy some amount of coin $\Delta \ge 0$ from the market. Such purchases
will cause some amount of slippage in the reported price $g$, causing some loss to the LP, which may be recouped with fees.
The question is, assuming that this CFMM is the only available market, what is the largest trade that a trader can perform
such that a liquidity provider still has positive payoff from the trade?

\paragraph{Curvature and profits.} Formally, suppose that an LP provides all of the assets $R, R' \in \reals_+$ to a CFMM with $\mu$-stable price impact function $g$. We assume that the CFMM charges some fee $(1-\gamma)$, but this fee is not included in $g$. The no-fee portfolio value of this CFMM is:
\[
g(0)R + R'.
\]
After a price change to $m_a = g(-\Delta) \le g(0)$, the opportunity cost (sometimes called the `impermanent loss') of this portfolio is given by
\[
\underbrace{(g(-\Delta)(R + \Delta) + R'- \Delta')}_\text{LP portfolio value} - \underbrace{(g(-\Delta)R + R')}_\text{Equivalent portfolio value} = g(-\Delta)\Delta - \Delta',
\]
and, by definition of marginal price
\[
\Delta' = \int_0^{-\Delta} g(t)\,dt.
\]
Since $g$ is a nonincreasing function, we have
\[
\Delta' = \int_0^{-\Delta} g(t)\,dt \le g(0)\Delta,
\]
so,
\[
g(-\Delta) \Delta - \Delta' \ge (g(-\Delta) - g(0))\Delta \ge -\mu \Delta^2,
\]
is a lower bound on the opportunity cost. Here, the second inequality follows from the definition of $\mu$-stability.

On the other hand, if $g$ is a marginal price function with some fee $0 < \gamma \le 1$, the value of fees earned is at least $(1-\gamma)\Delta g(-\Delta) = (1-\gamma)\Delta m_a$, where $m_a$ is the new price (see appendix~\ref{app:pv-lower-bound} for a general statement and proof), so LPs are guaranteed to make a profit whenever
\[
(1-\gamma)\Delta m_a > \mu\Delta^2,
\]
which happens when
\begin{equation}\label{eq:trade-bound}
\Delta < \frac{(1-\gamma)m_a}{\mu}.
\end{equation}
\paragraph{Discussion.} This inequality shows that a sufficient condition on the trade size for which LPs still make a profit is inversely proportional to the curvature bound on the CFMM.
Additionally, using this formula, we can compute a lower bound on the rate of growth of profits for LPs as a function of fees and curvature, given a distribution of trades, $\Prob[\Delta \leq x]$. An extension of this result can provide a discrete time, curvature-based analogue of~\cite{tassy_white} that generalizes to a number of CFMMs other than Uniswap.

Another interpretation of inequality~\eqref{eq:trade-bound} is that, as the effective curvature decreases, traders can perform large trades, while liquidity providers still come out ahead, relative to an equivalent portfolio which simply
holds $R$ of the traded coin and $R'$ of the num\'eraire. Note that this inequality comes from the fact that the trade
does not depend on the future price of the coin, which we will call an `uninformed' trade, and such trades are, as discussed previously, very common in stablecoin--stablecoin trades.

\subsection{Informed trading}\label{sec:informed}
On the other hand, liquidity provider losses change drastically
when we have an agent who attempts to maximize their profits given information about future prices.
In order to model this phenomenon, we need to describe a participant other than the LP, who has some amount
of knowledge of future prices.
Analogous to~\cite{aoyagi2020lazy} and the classical market microstructure models~\cite{kyle1985continuous, glosten1985bid}, we will consider a market with an LP and an informed trader under the assumption of no-arbitrage.
We will construct a two-player game between an informed trader who can predict the next price update of some external market with non-trivial edge, and a liquidity provider whose funds are locked in the CFMM.
Using this game, we will show a profit (or loss) lower bound for both LPs and informed traders, akin to those used to describe market maker profits in open limit order books~\cite{glosten1985bid, glosten1994electronic}.
From this lower bound, we will show that informed traders need less of an informational edge to guarantee that trading with a lower curvature CFMM has profits as large as trading with a higher curvature CFMM.

\paragraph{Problem set up.} In this game, as before, we have two agents: a liquidity provider and
an informed trader, where the informed trader is allowed to trade with the CFMM.

We will assume that the CFMM, with fee-less marginal price function $g$ (such that the marginal price with fees is $g^f(-\Delta) = \gamma g(-\gamma \Delta)$ with $\Delta \ge 0$) and the reference market both start at some fixed price $m_0 = \gamma g(0)$. We will assume
the function $g$ is $\mu$-stable and $\kappa$-liquid in some interval $0 \le \Delta \le L$. The informed trader then knows that the reference market price will decrease to some amount $m_1 \le m_0$ with probability $\alpha$ or stay at $m_0$ with probability $(1-\alpha)$. By no-arbitrage, any price discrepancies between the CFMM and the reference market are immediately removed, so the informed trader must make a trade which maximizes the expected profit, before the trader is able to see the new price.

\paragraph{Informed trader edge.} The expected edge of an informed trader under this framework is given by
\[
E_V(\Delta) = \underbrace{\int_0^{\Delta} \gamma g(-\gamma t) \,dt}_\text{total profit for trading $\Delta$} - \underbrace{(\alpha m_1\Delta + (1-\alpha)m_0\Delta)}_\text{expected profit for holding $\Delta$ asset}.
\]
We can rewrite this in a slightly simpler form by noting that, by assumption, $m_0 = \gamma g(0)$, so
\[
E_V(\Delta) = \gamma\int_0^{\Delta} (g(-\gamma t) - g(0))\,dt + \alpha(m_0 - m_1)\Delta.
\]
Using the fact that $g$ is $\mu$-stable, we have $g(-\gamma t) - g(0) \ge -\mu\gamma t$, and that
\[
E_V(\Delta) \ge \alpha(m_0 - m_1)\Delta - \frac12 \mu\gamma^2\Delta^2.
\]
Taking the supremum of both sides over $\Delta \ge 0$ (since an informed traders seek to maximize
their profit) gives that
\[
E_V^\star \ge \frac{\alpha^2(m_0 - m_1)^2}{2\mu\gamma^2}.
\]
Here, the surprising fact that the fee $\gamma$ appears in the denominator (versus, as one might expect, in the numerator) happens because
the price $m_0 = \gamma g(0)$ depends implicitly on the fee. This result holds independent of the interval size $L$ if $g$ is $\mu$-stable for all $\Delta \ge 0$. Note that, if $\mu$ is very small (\ie, the CFMM has low curvature) then $E_V^\star$ is large; similarly, given two CFMMs, one with lower and one with higher curvature, $\alpha$, the edge, needs to be larger in the CFMM with higher curvature to achieve the same lower bound for the payoff.

\paragraph{Liquidity provider loss.} We can similarly get a lower bound on the expected loss of an LP since it is equal to
$-E_V(\Delta)$:
\[
-E_V(\Delta) = \gamma\int_0^{\Delta} (g(0) - g(-\gamma t))\,dt - \alpha(m_0 - m_1)\Delta,
\]
but, if $\Delta \le L$, we have that $g(0) - g(-\gamma t) \ge \gamma \kappa t$, and so
\[
-E_V(\Delta) \ge \frac12 \kappa\gamma^2 \Delta^2 - \alpha(m_0 - m_1)\Delta.
\]
Minimizing both sides gives
\begin{equation}\label{eq:adverse}
-E_V^\star \ge -\frac{\alpha^2(m_0 - m_1)^2}{2\kappa \gamma^2}
\end{equation}
whenever $\alpha(m_0 - m_1) \le L\kappa \gamma^2$ (\ie, when the unconstrained minimum lies in the interior of
the interval $[0, L]$) and is otherwise bounded by $-E^\star_V \ge \kappa\gamma^2 L^2/2 - \alpha(m_0 - m_1)L$.
We will mostly consider the first case in the following discussion, since we can often expand the interval
$L$ to be large enough to contain this bound. (Note also that~\eqref{eq:adverse} is equivalent to giving an
upper bound on the expected value of an informed trader.)

\paragraph{Discussion.} This matches the empirical observation that lower curvature CFMMs tend to have higher liquidity for assets that do not require much information to trade whereas LPs of higher-curvature CFMMs lose less to informed traders. This result illustrates that unlike common wisdom in the CFMM design space, one need not only have an optimal fee to maximize LP returns, but one needs to adjust the curvature as well.

\begin{figure}
    \centering
    \includegraphics[width=.6\textwidth]{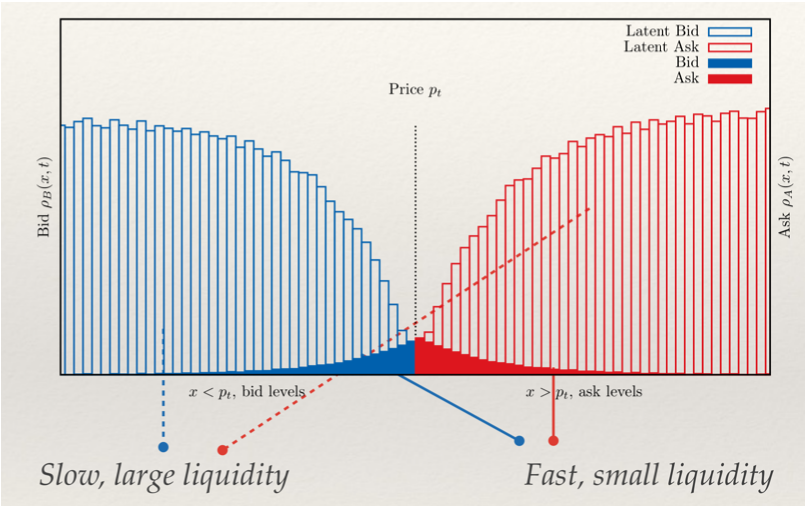}
    \caption{Order book shapes for different liquidity levels. Glosten and others have show that the concave shape is better for large liquidity and low trading velocity (\eg, trades per second) whereas the convex shape is better for many small orders and high trading velocity. Figure from~\cite{lemhadri2018market}.}
    \label{fig:shape_order_book}
\end{figure}

Moreover, this result represents an analogue of classical microstructure results that show that the shape of an order book gives bounds on adverse selection.
Glosten~\cite{glosten1994electronic} showed that when you consider market makers who have to quote prices on multiple markets, then the shape of the order book impacts how liquidity changes in response to adverse selection.
In figure~\ref{fig:shape_order_book}, we see two different shapes for an order book, one approximately concave (the unshaded bars) and one convex (the filled in region).
When a market maker observes or realizes adverse selection costs, they make a market more illiquid (\eg, by canceling orders) to force active traders to pay a higher impact cost to market makers.
This leads to the higher curvature, concave shape seen in figure~\ref{fig:shape_order_book}.
Equation~\eqref{eq:adverse} then suggests that market makers in CFMMs can replicate the same effect by increasing the curvature, therefore increasing the curvature lower bound, $\kappa$.

\section{Price stability and yield farming}\label{sec:price-stability}
In this section, we will describe price stability when arbitrageurs trade between two markets, each with different curvature bounds.
This stability result provides a quantitative explanation for the stability phenomenon of figure~\ref{fig:susd}.
Given that the empirically observed sUSD price instability was due to liquidity incentives (yield farming), it is natural to expect that there is a relationship between curvature and the precise costs of a stability incentive.
In other words, the question we seek to answer is: how much do we need to pay LPs for providing liquidity?
We describe this connection precisely in~\S\ref{sec:yield_farming_returns}, which shows that there is an optimal liquidity subsidy
for a market that interacts with an external market with different curvature.

\subsection{Model description}\label{sec:model-description}
Here, we will define the market model used throughout the remainder of this section. In our model, we have two available markets:
the external market (whose price fluctuates due to extrinsic demand) and the secondary market (which we will assume is a CFMM, though
the model holds more generally),
along with an arbitrageur agent which seeks to maximize their profit by exploiting the difference in price between these two
markets.

\paragraph{Market model.}
We describe a relatively simple, but very general, model of the \emph{external market} and how it interacts with the given CFMM. In particular,
the external market reports some strictly positive price $m_0 \in \reals_{++}$ at the start of the round. The basic model of interactions between the markets and the arbitrageur proceeds as follows:
\begin{enumerate}
	\item At the round start, the quoted external market price is $m_0^e$, while the secondary market price is $m_0^s$.
	\item An arbitrageur then trades with the external market and the secondary market (which will usually be a CFMM). This results in a new external \emph{and} secondary market price $m_a$ which are equal since no-arbitrage has been enforced.
	\item The external market price then changes from the no-arbitrage price $m_a$ to a new price by some process modeling external influences. Step 1 is repeated with the new external and secondary market prices.
\end{enumerate}
In fact, in our presentation, we will not assume anything about the dependence of the new price on $m_a$ or even on $m_0^e$ or $m_0^s$,
which means that the results here hold for essentially all (reasonable) models of exogenous price changes for the external market price.

\paragraph{Main goal.} We will show that, even when the external market price $m_0^e$ differs widely from the secondary market price $m_0^s$, the (new) arbitraged market price $m_a$ does not differ too much from the previous secondary market price $m_0^s$. Written out, we wish to find conditions such that, even when the price difference between both markets before no-arbitrage, $m_0^e - m_0^s$, is large, the difference between the no-arbitrage price and the secondary market's price $m_a - m_0^s$ is small, in a precise sense. This would imply that even though the external market price deviates from the secondary market price $m_0^e - m_0^s$, the secondary market is able to force the new no-arbitrage price, $m_a$, back to a price that was close to its previous value, $m_0^s$.

\paragraph{Assumptions.} In this set up, the external market will have a price impact function $f$ which is, as before, continuous and nondecreasing. We will define the initial price of the external market as $m_0^e = f(0)$. Additionally, we will assume the external market is $\kappa$-liquid, with a slightly different definition than the one given in~\S\ref{sec:cfmm-curvature}: we will say an external market is $\kappa$-liquid if it satisfies, for $\Delta \ge 0$,
\[
f(\Delta) - f(0) \ge \kappa \Delta.
\]
This differs from the original definition given in~\S\ref{sec:cfmm-curvature} because, here, $\Delta$ is the amount \emph{purchased} from the external market, rather than the amount sold to it. In this case, if $f$ is a differentiable convex function, we have that $\kappa = f'(0)$ is the tightest possible constant $\kappa$ satisfying this condition and is a global bound (holding for all $\Delta \ge 0$) which follows immediately from the first-order convexity conditions.

As before, we will simply assume that the secondary market, with continuous, nondecreasing price impact function $g$, is $\mu$-stable with the usual definition given in~\S\ref{sec:cfmm-curvature}. We will similarly define $m_0^s = g(0)$.


\subsection{Stability and Curvature}
In this section, we derive the main result for general, continuous price impact functions, satisfying the conditions outlined previously. 


\paragraph{Main result.} Assume the price impact function of the primary market, $f$, is $\kappa$-liquid (in the sense above) and that the price price impact function of the secondary market, $g$, is $\mu$-stable. We will show that the secondary-market's price change is bounded in the following way:
\begin{equation}\label{eq:main}
m_0^s - m_a \le \frac{\mu}{\kappa}(m_0^s - m_0^e).
\end{equation}
Note that because both sides of the inequality are nonnegative, inequality~\eqref{eq:main} can also be written as
\[
|m_0^s - m_a| \le \frac{\mu}{\kappa}|m_0^s - m_0^e|.
\]
In other words, the no-arbitrage price
change is at most a factor of $\mu/\kappa$ from the difference between the primary and secondary markets. This quantity (and therefore the price change after arbitrage) is small whenever the secondary market is very liquid ($\mu$ is small), or when the external market is very illiquid ($\kappa$ is large). While apparently simple, we show in~\S\ref{sec:yield_farming_returns} that this result can be applied to many useful circumstances.

\paragraph{Proof of main result.} By assumption, we have
\[
f(0) = m_0^e \le m_0^s = g(0),
\]
so $f(0) \le g(0)$. Then, if we can find any $\Delta \ge 0$ that satisfies
\begin{equation}\label{eq:continuity-bound}
f(\Delta) \ge g(-\Delta),
\end{equation}
then there exists some $0 \le \Delta^\star \le \Delta$ such that
\[
f(\Delta^\star) = g(-\Delta^\star),
\]
by continuity and monotonicity of $f$ and $g$. As before, the no arbitrage price $m_a$ is defined
as $m_a = f(\Delta^\star) = g(-\Delta^\star)$.

To show that there exists a $\Delta$ satisifying~\eqref{eq:continuity-bound}, note that any $\Delta \ge 0$ which satisfies
\begin{equation}\label{eq:assumption}
m_0^e + \kappa \Delta \ge m_0^s,
\end{equation}
automatically satisfies~\eqref{eq:continuity-bound} since
\[
f(\Delta) \ge f(0) + \kappa \Delta = m_0^e + \kappa \Delta \ge m_0^s = g(0) \ge g(-\Delta)
\]
where the first inequality follows from~\eqref{eq:k-liquidity}, the second inequality follows from~\eqref{eq:assumption}, while the last inequality follows from the fact that $g$ is a nondecreasing function.

In order to satisfy~\eqref{eq:assumption}, we can easily choose
\[
\Delta = \frac{m_0^s - m_0^e}{\kappa}.
\]
Such a $\Delta$ will then satisfy~\eqref{eq:continuity-bound}. This, in turn, implies that a no-arbitrage trade $\Delta^\star$ satisfies $0 \le \Delta^\star \le \Delta$, and
\[
m_0^s - m_a = g(0) - g(-\Delta^\star)
\le g(0) - g(-\Delta) \le \mu \Delta = \frac{\mu}{\kappa}(m_0^s - m_0^e).
\]
Here, the first equality follows from the definition of $m_0^s$ and $g$, while the first inequality follows from the monotonicity of $g$ and the second inequality follows from~\eqref{eq:mu-stability}. The resulting inequality is the one given in~\eqref{eq:main}.

\begin{figure}
    \centering
    \includegraphics[scale=0.6  ]{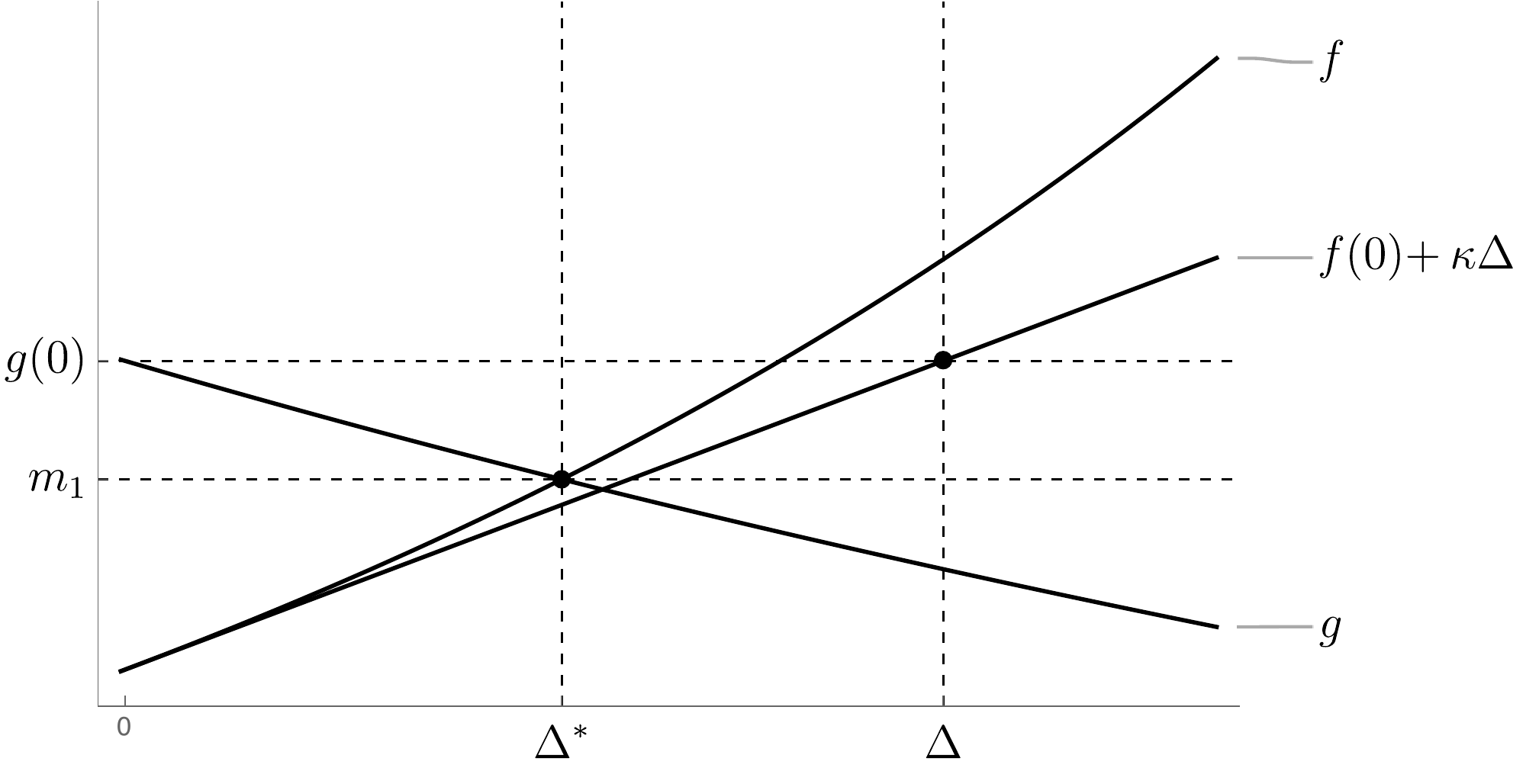}
    \caption{Visual proof of the main result. Given $g(0)>f(0)$ and $f$ and $g$ monotonic and continuous, we find $\Delta$ such that $f(\Delta) \ge g(-\Delta)$. The fact that there exists $0 \le \Delta^\star \le \Delta$ such that
$f(\Delta^\star) = g(-\Delta^\star)$ is then obvious from the figure. We also note that $\Delta=\frac{g(0)-f(0)}{\kappa}$ will be a worse overestimator for $\Delta^\star$ if the curvature of the two impact functions is large.}
    \label{fig:proof}
\end{figure}

\begin{figure}
    \centering
    \includegraphics[scale=0.35]{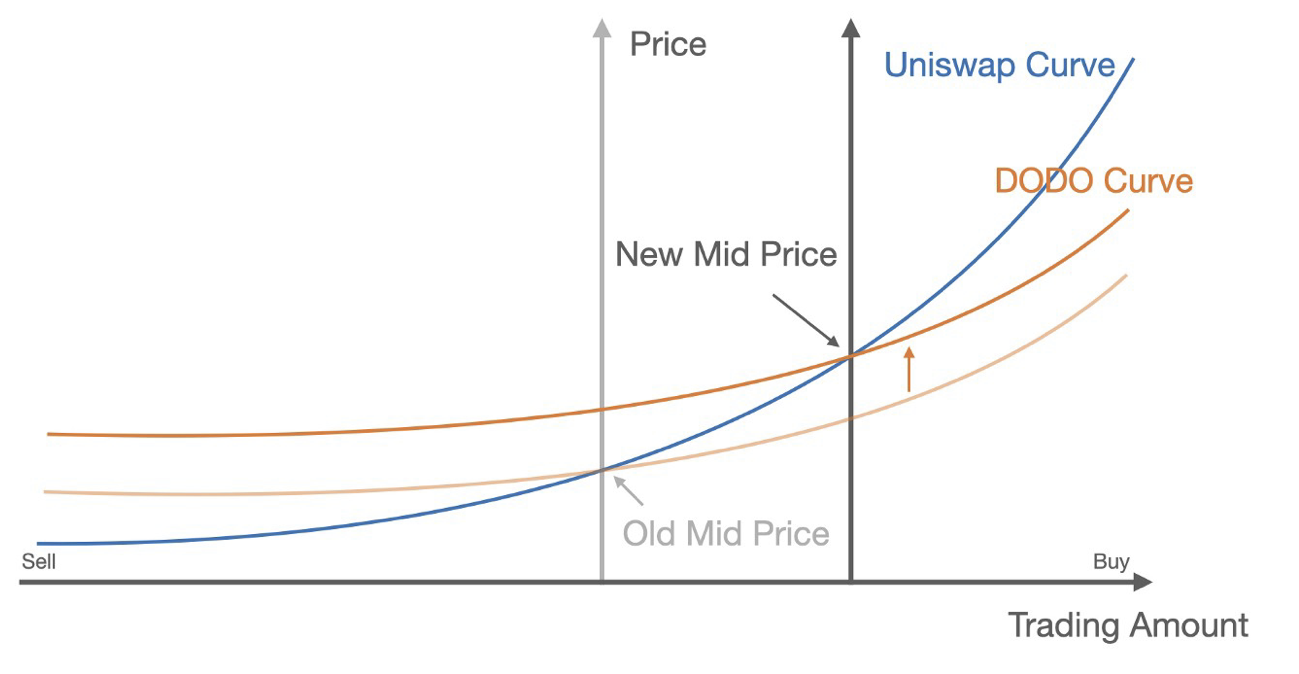}
    \includegraphics[scale=0.18]{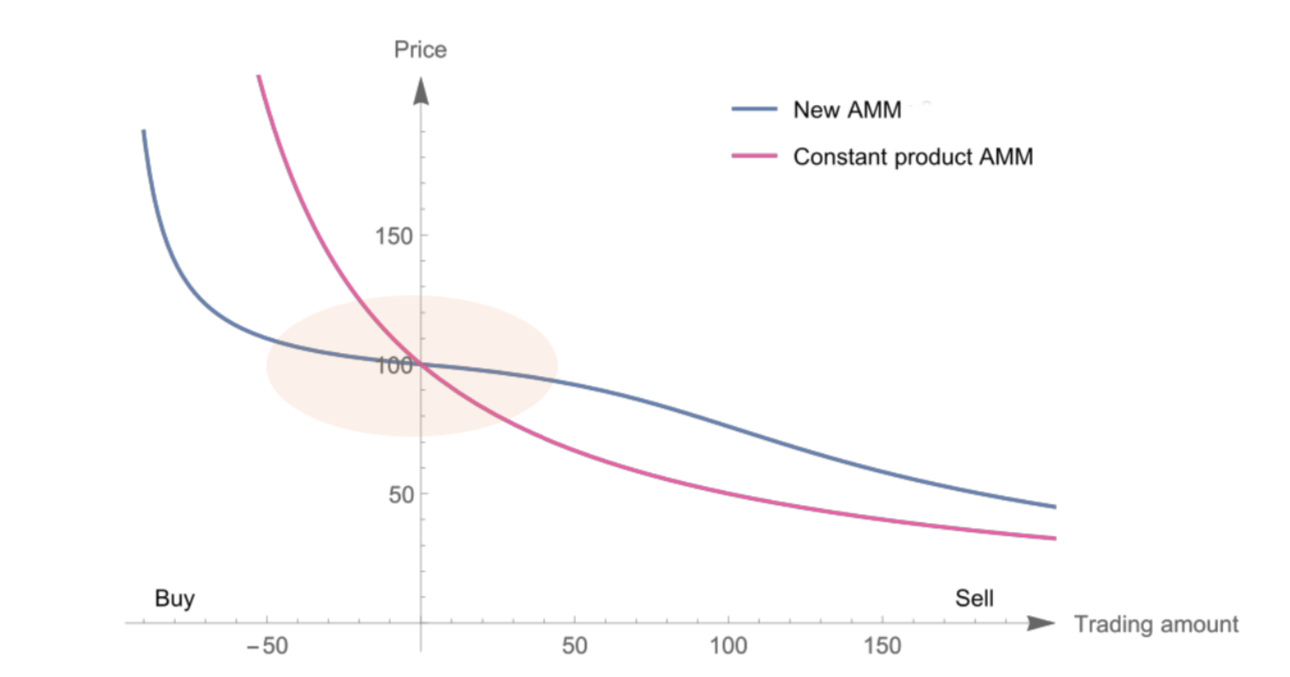}
    \caption{Marginal price curves for Dodo~\cite{dodo_curve} (left) and McDEX~\cite{mcdex}. Note that Dodo's curve flattens in response to a trade, whereas McDEX's curve has asymmetric slippage. Both of these curves aim to provide leverage, so they have higher curvature when the protocol takes more risk (\eg, McDEX when the protocol's net position is negative). For the McDEX curve, we can view this as $\mu^- > \mu^+$ for some interval $[-50, 50]$}
    \label{fig:asymmetric-curves}
\end{figure}

\begin{figure}
    \centering
    \includegraphics[width=.95\textwidth]{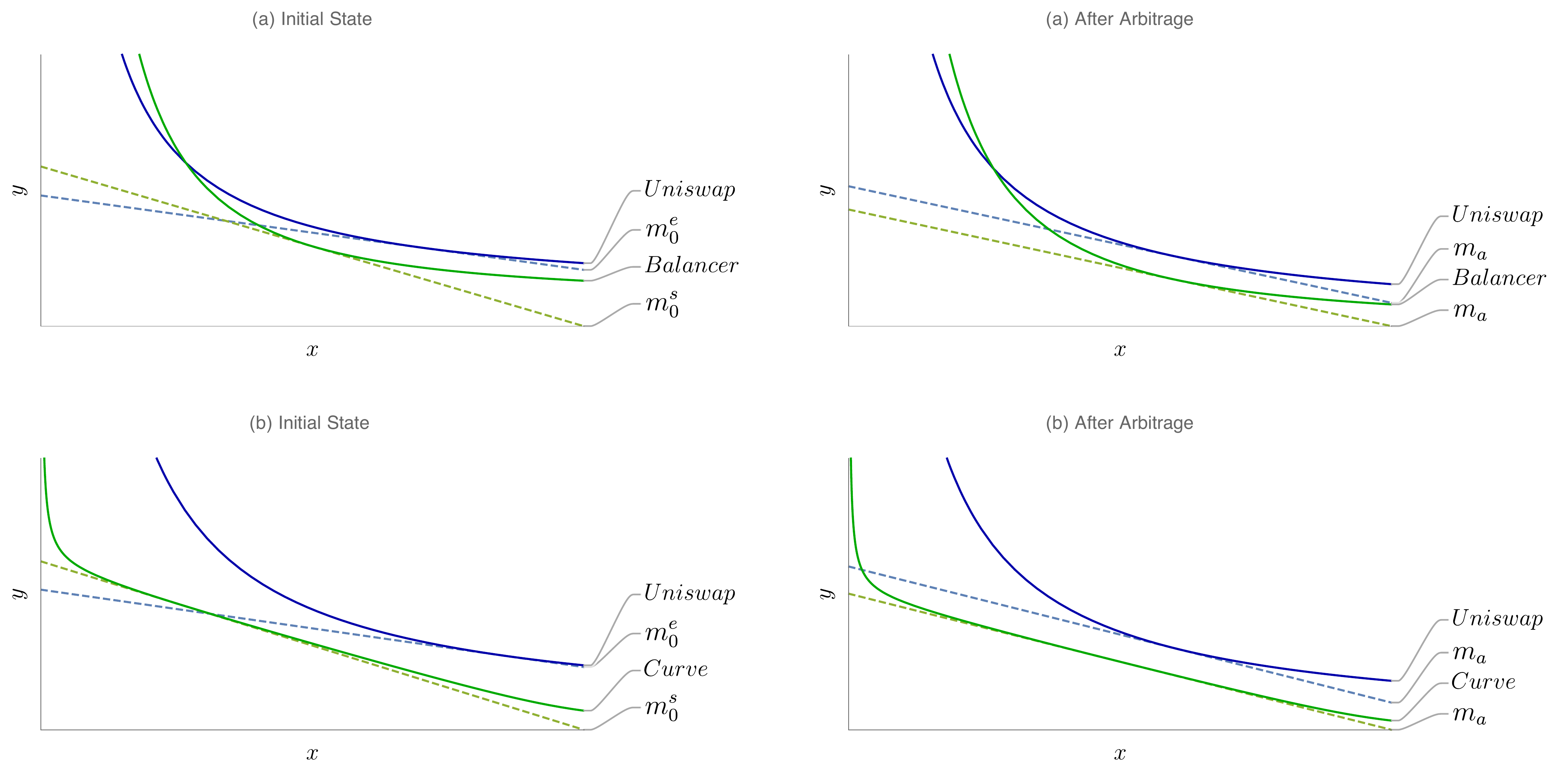}
    \caption{Visual illustration of arbitrage between two CFMMs with different curvature. Note the changes in slope, corresponding to the changes in market price after no-arbitrage is enforced.} 
    
    \label{fig:arb_plots}
\end{figure}

\paragraph{Assumptions.} While the assumption that $m_0^e \le m_0^s$ might appear restrictive, it is actually
fully general. For example, if $f(\Delta)$ and $g(-\Delta)$ specify the price of an asset $A$ with respect to a tradeable asset $B$,
after buying $\Delta$ amount of asset $A$ from the primary (or secondary) market, then the amount of asset $B$ traded with each
market is given by the quantity functions
\[
p(\Delta) = \int_0^\Delta f(t)\,dt, \qquad q(-\Delta) = \int_0^\Delta g(-t)\,dt.
\]
On the other hand, we may ask what the price of asset $B$ is with respect to asset $A$ after buying some amount $\Delta'$ of
asset $B$. The quantity of asset $A$ received for $\Delta'$ of asset $B$ is easily seen to be $p^{-1}(\Delta')$ for the primary
market and $q^{-1}(\Delta')$ for the secondary market. Both exist because
$f$ and $g$ are strictly positive which implies that $p$ and $q$ are strictly monotonic. This implies
that the respective price (using implicit differentiation) is given by
\[
(p^{-1}(\Delta'))' = \frac{1}{p'(p^{-1}(\Delta'))} = \frac{1}{f(p^{-1}(\Delta'))} = \frac{1}{f(\Delta)},
\]
where we have defined $\Delta = p^{-1}(\Delta')$, and similarly for $q$. So, if $m_0^e \ge m_0^s$, we may always
`swap' asset $A$ for asset $B$ in this sense, such that the resulting marginal prices are given by $1/m_0^e \le 1/m_0^s$,
and enforce no-arbitrage conditions over coin $B$, instead.

\paragraph{Extensions.} As with~\S\ref{sec:cfmm-curvature}, may not be the case that constants $\mu$ or $\kappa$ exist for trades of all possible sizes. In general, such constants do exist for trades of bounded size, say $0 \le \Delta \le L$. In this case, the main result extends immediately in the following way: for
any prices satisfying
\begin{equation}\label{eq:condition}
m_0^s - m_0^e \le \kappa L,
\end{equation}
we have
\[
m_0^s - m_a \le \frac{\mu}{\kappa}(m_0^s - m_0^e).
\]
The proof of this statement is identical to the one above, with the additional condition that the primary
and secondary market prices differ by no more than $\kappa L$.

\subsection{Yield farming subsidy}\label{sec:yield_farming_returns}
One of the main drivers of the growth in CFMM usage in 2020 was \emph{yield farming}.
Yield farming, which is similar to maker--taker rebates in traditional trading~\cite{brolley2013informed}, involves subsidizing the provision of liquidity for a new issued crypto asset.
Suppose that some asset is issued at time $t_0$ by a smart contract and that it has an inflation schedule $i_t \in \reals_+$, where $i_t$ is the number of units of X produced at time $t$.
In order to incentivize liquidity between the new asset and a num\'eraire, the smart contract reserves some percent (say, $\ell_t$) of inflation for liquidity provision.
If a user creates LP shares for a CFMM, which trades this coin and then stakes (or, in other words, locks) these shares into a smart contract, then they receive some amount of this coin from the smart contract for providing liquidity.
For instance, if there are 1000 LP shares locked into the contract and a single user created 100 of these locked shares, they might receive $\frac{100}{1000}i_t\ell_t$ units of the new asset at time $t$.
By subsidizing liquidity, the smart contract issuing the new coin can ensure that users can trade the new asset while also ensuring that liquidity providers have lower losses. The main loss that CFMM liquidity providers face is `impermanent loss' or losses due to the concavity of the portfolio value of an LP share; see, \eg,~\cite{clark2020replicating} for the specific case of Uniswap. One can also directly show these losses occur by using the definition of the portfolio value in~\cite{angeris2020improved}.


\paragraph{Sufficient subsidy.} A protocol designer would then want
to ensure that LPs are compensated enough, say, by some amount $R_\ell$ in the traded coin, to
have nonnegative profit after a no-arbitrage trade with the external market; \ie, we need to guarantee that the portfolio value of an LP, after arbitrage and subsidy, is nonnegative. To
do this, note that the opportunity cost or `impermament loss' of being an LP in the secondary market is given by (in a similar way to~\S\ref{sec:mean-revert}):
\begin{align*}
m_a\Delta - \int_0^{-\Delta} g(-t)\,dt \ge (m_a - g(0))\Delta = (m_a - m_0^s)\Delta,
\end{align*}
where we have used the fact that $g$ is nondecreasing, and $g(0) = m_0^s$ by definition. Applying~\eqref{eq:main} gives
\begin{equation}\label{eq:il_subsidy}
(m_a - m_0^s)\Delta \ge - \frac{\mu}{\kappa}(m_0^s - m_0^e).\end{equation}
Therefore, to incentivize LPs to continue adding liquidity to the $\mu$-stable secondary market, assuming an external market that is
$\kappa$-liquid, it suffices to subsidize the LPs by some amount
\[
R_\ell' = \frac{\mu}{\kappa}(m_0^s - m_0^e),
\]
in the num\'eraire. Alternatively, they can be subsidized by at least
\[
\frac{R_\ell'}{m_a} = \frac{\mu}{\kappa m_a}(m_0^s - m_0^e) \le \frac{\mu}{\kappa}\left(1 - \frac{m_0^e}{m_0^s}\right)
\]
in the traded coin, since $m_0^e \le m_a \le m_0^s$. In other words, the total quantity of subsidy is proportional to the curvature
of the primary market, and inversely proportional to the curvature of the secondary market. If we define $h = m_0^e/m_0^s$ to
be the percentage growth of the asset (note that $h \le 1$ since $m_0^e \le 	m_0^s$) then we have simple expression for the amount of
subsidy that is sufficient, in the traded asset, which we will define as:
\begin{equation}\label{eq:il_growth}
R_\ell = \frac{\mu}{\kappa}(1 - h).
\end{equation}

\paragraph{Discussion.} This gives a simple condition which guarantees that liquidity providers
have nonnegative returns for providing liquidity. In particular, we note that more subsidy has to be provided
as $h$ becomes small (\ie, the price is changing with large drift) or when $\mu/\kappa$ is large (\ie,
the secondary market, for which the LPs are providing liquidity for, is very illiquid when compared to the external market).
In general, this means that how much subsidy one might need to provide depends not just on
the drift of the asset, but also the relative curvature of the two markets.

\section{Dynamic hedging of CFMMs}\label{sec:dynamic_hedging}
We describe how dynamic hedging quantities for contingent claims (\eg, an asset's Delta and Gamma) are constructed for CFMMs, providing a means for LPs to hedge risk such as so-called `impermanent loss.'
In order to simplify notation, we will refer to the Delta and Gamma of a portfolio as $P_{\Delta}$ and $P_{\Gamma}$, respectively.
To connect portfolio value to curvature, we will show that analogues of dynamic hedging Greeks are closely related to curvature.
These quantities are important for liquidity providers as they represent the net exposure to the underlying collateral the LPs have, as well as how to hedge loss due to drift (known colloquially in decentralized finance as `impermanent loss').
While Uniswap can be statically replicated by a portfolio of options~\cite{clark2020replicating}, it is unclear precisely how to do this for generic CFMMs. We will compute dynamic hedging quantities for LPs here and show how they behave under trades.
This behavior, which will be connected to curvature, represents extra convexity in the replicating portfolio that is needed to compensate for the LP share of a high curvature CFMM.

\paragraph{Approximate hedges.} We apply these hedging results to try to construct approximate hedges for impermanent loss.
If the price of the traded coin increases while volatility is mollified, the LP realizes impermanent loss~\cite{angeris2019analysis, tassy_white}.
As the price drifts up, if the LP is able to increasingly sell put options for the tradable asset, then they can lower their realized impermanent loss when they exit the pool.
The question then is: how many put options does an LP need to sell as a function of the price upon their entry intro the pool?

We illustrate that if options on realized impact costs exist, then an LP can enter a short put option portfolio on impact costs to hedge their impermanent loss.
Such options correspond to options on the change in the market price when a trade of size $\Delta$ is made.
Owners of such an option effectively have insurance on price changes in the underlying due to a single large trade.

\paragraph{Delta hedging in practice.} While unnatural relative to conventional options on stocks, options on impact cost exist in traditional finance when considering American Depository Receipts (ADRs) in the US equity market.
Suppose that there is a stock that exists both in a non-US market and on a US exchange as an ADR.
ADRs allow for the non-US stock to be traded in the US via a synthetic asset that has a creation and redemption mechanism.
When the foreign exchange rate between the non-US currency and US dollars drifts wildly, there is an impact of foreign exchange rate on option pricing.
The difference in price between options on the non-US stock and options on US ADRs replicates an option on the impact cost of the foreign exchange rate on the stock price~\cite{madan2012jointly, maltritz2010currency, anderluhadr}.

\subsection{Computing the Greeks}
The portfolio value of the reserves, as before, is defined as
\[
P_V = mR + R',
\]
where $R$ and $R'$ depend implicitly on the price $m$ and the value of $\psi(0, 0)$. We then have:
\begin{equation}\label{eq:delta-gamma-defs}
P_\Delta = \frac{dP_V}{dm} = R, \qquad P_\Gamma = \frac{d^2P_V}{dm^2} = \frac{dR}{dm},
\end{equation}
which we show below. (Here $P_\Delta$ and $P_\Gamma$ represent the corresponding Greeks of $P_V$.)

\paragraph{Computation.} Because it is often the case that the function $\psi$ is of the following
form (when there are no fees) 
\[
\psi(R, R', \Delta, \Delta') = \Psi(R - \Delta, R' + \Delta'),
\]
for some function $\Psi: \reals^2 \to \reals$ (as is the case with all examples presented in~\S\ref{sec:cfmm-curvature}),
it suffices to consider the function $\Psi$ in terms of only the reserves $R$ and $R'$. We will assume this
is true in the following derivation.

As before, the portfolio value is given by
\[
P_V = mR + R',
\]
where $m$ is the market price, which, by no arbitrage, must satisfy (at the reserve values $R$ and $R'$, after no arbitrage)
\[
m = \frac{\partial_1\Psi(R, R')}{\partial_2\Psi(R, R')}.
\]
(Note that there is no negative sign here, due to the definition of $\Psi$.)
By implicitly differentiating $\Psi(R, R')$ over any level set, we have that
\[
\frac{d\Psi(R, R')}{dm} = (\partial_1\Psi(R, R')) \frac{dR}{dm} + (\partial_2\Psi(R, R'))\frac{dR'}{dm} = 0,
\]
so,
\[
m\frac{dR}{dm} = - \frac{dR'}{dm}.
\]
Using this final condition:
\[
P_\Delta = \frac{dP_V}{dm} = R + m\frac{dR}{dm} + \frac{dR'}{dm} = R,
\]
as required, while the expression for $P_\Gamma$ follows from the fact that
\[
P_\Gamma = \frac{dP_\Delta}{dm} = \frac{dR}{dm}.
\]

\paragraph{Approximate hedging.} On the other hand, the deterministic price schedule of CFMMs also implies that directly hedging trade quantities is equivalent to hedging prices if the CFMM is the primary market. In the case that $g$ is convex, which we assume here, we can give bounds that can be easily replicated with options. We note that this result only holds
for decreasing prices, due to the assumptions on $\mu$ and $\kappa$, but we suspect there exist more general bounds, holding over
any change in price, which are very similar in spirit.

Suppose that an LP has reserves $R$ and $R'$ in a CFMM and the CFMM is the primary market.
First, note that,
\[
R = P_\Delta = \frac{dP_V}{dm} = \frac{d P_V}{d \Delta} \frac{d\Delta}{dm}.
\]
After a trade of size $\Delta$, we have
\[
P_V(-\Delta) = g(-\Delta)(R + \Delta) + (R' - \Delta'),
\]
so that if $g$ is differentiable, we have
\begin{equation}\label{eq:pv_deriv_alpha}
    \frac{d P_V}{d \Delta} = g'(-\Delta) (R + \Delta) + g(\Delta) - \frac{d\Delta'}{d\Delta} = g'(-\Delta) (R + \Delta), 
\end{equation}
where we used \eqref{eq:g_def} in the last equality.
Therefore, if our CFMM is $\mu$-stable in the sense of \eqref{eq:mu-stability}, then, because $g$ is convex, nondecreasing, we have that $g'(-\Delta) \le \mu$, so
\[
\frac{dP_V}{d\Delta} = g'(-\Delta) (R + \Delta) \leq \mu(R + \Delta).
\]
Finally, because $\frac{d\Delta}{dm} \le 0$, then we have that
\[
P_\Delta = \frac{dP_V}{d\Delta} \frac{d\Delta}{dm} \ge \mu (R+\Delta)\frac{d\Delta}{dm}.
\]
If the CFMM is both $\mu$-stable and $\kappa$-liquid, then we have the following expansion:
\begin{equation}\label{eq:delta_hedge}
\mu (R+\Delta)\frac{d\Delta}{dm} \le P_\Delta \le \kappa(R+\Delta)\frac{d\Delta}{dm}.
\end{equation}
These inequalities show that the curvature constants provide means for super and subhedging of LP share risk.
In particular, these linear bounds allow for an LP to hedge their risk using simpler instruments, as described in the sequel.
Finally, we note that using these calculations, one can show that hedging a portfolio of $n$ stablecoins is approximately equivalent to hedging two stablecoins via an $n$-dimensional generalization presented in Appendix \ref{app:ncoinpv}.

\section{Conclusion and future work}
In this paper, we explored how the shape of a constant function market maker affects its ability to serve as the primary market for digital assets.
To do this, we defined a notion of curvature for a given market and then showed its implications for liquidity providers in the market, showing that this notion of curvature is very closely related to the intuitive notion of curvature in the case of constant function market makers.

\paragraph{Liquidity provider returns.} There has been much empirical evidence suggesting that the return profile of a CFMM liquidity provider depends on the shape of the CFMM's trading function.
We studied the LP returns under three different scenarios: uninformed trading, informed trading, and yield farming.
Some of the results presented take inspiration from the traditional market microstructure literature and consider the return profile of informed traders.
These traders can be viewed as bringing information to the market by placing a Kelly-style bet on the next price update that will take place in the CFMM.
We were able to use no-arbitrage to derive lower bounds on the liquidity provider loss (and, conversely, lower bounds on the edge trader's expected profit) akin to that of Glosten and Milgrom which connect curvature to the informed trader's informational edge. The results from this section provide a simple economic interpretation of curvature as the amount of information an informed trader needs to achieve a certain profit (given a fixed edge and market price).

\paragraph{Yield farming.}
We then extended the definition to also include interacting markets with finite liquidity.
These notions of shape or curvature could then be used to capture how a single trade on a market with finite liquidity affects prices on another market with finite liquidity.
Using these definitions for curvature, we were able to bound the tracking error when an arbitrageur trades between a pair of markets with different curvatures.
When specialized to CFMMs, the results presented generalize some of the results of~\cite{angeris2019analysis, angeris2020improved} to the case where the market is not infinitely liquid.

We then used this to analyze the yield farming phenomenon, where protocols began providing subsidies to liquidity providers. Here, we showed a lower bound to the amount of subsidy needed to pay liquidity providers to account for their `impermanent loss' when compared to a market with bounded liquidity, which depended on the curvature of both markets and the rate of growth of the asset.
Combined, these results suggest that the curvature of a CFMM needs to be optimized to avoid adverse selection while also capturing trading volume and fees related to asset price growth.

\paragraph{Future work.}
This work can be extended in a number of ways.
On the practical side, many of the results presented here only work in two dimensions (\eg, two asset trading).
Generalizing our results to $n$ dimensions would be a useful but potentially difficult problem. For example, it is not clear how to define $\mu$-stability in higher dimensions for general CFMMs, without being overly restrictive.
Moreover, even though we give some sufficient conditions on the curvature of a `good' CFMM for certain applications, it is still an open question for how to take a given price process, represented, say, as an It\^o process or a jump-diffusion process, and then construct a `good' CFMM.
If this were found, then one could take historical data for a crypto asset and construct an optimized CFMM for trading this asset.
Finally, it is clear that dynamic CFMMs~\cite{evans2020liquidity, zhang_2020, breeder_2020, niemerg2020yieldspace, notional}, \ie, CFMMs whose trading functions vary in time according to either a stochastic or control mechanism, continuously affect the curvature of the trading function.
Given the results of this paper, a natural extension to inquire about is: how should one design an optimal control mechanism to replicate a desired payoff or behavior?
    The results of this paper suggest that the trade-off between adverse selection and payoff growth are extremely important to such designs, especially for products with sharp payoffs or time decay (\eg, barrier options).

The results of~\S\ref{sec:yield_farming_returns} intimate that there is such a mapping, akin to super replication results from traditional mathematical finance.
Curiously, results from the mathematical finance literature also heavily use convex duality and it is likely that there are a number of fruitful translational results can be found.
These results will likely give more power to the dynamic hedging results of~\S\ref{sec:dynamic_hedging} and will be useful for comparing CFMM payoffs to traditional derivatives pricing.
We suspect that some of our conjectures in appendix~\ref{app:superreplication}, regarding the superhedging of contingent claims specified by CFMM portfolio values, is likely to be a problem with deep connections to such results.

\section*{Acknowledgments}
The authors would like to thank Yi Sun, Ciamac Moallemi, Hsien-Tang Kao, Victor Xu, Rei Chiang, and Adam Lerer for helpful comments and suggestions. We also would like to thank Matteo Liebowitz of Uniswap for providing data.

\bibliographystyle{amsplain}
\bibliography{bib}

\clearpage

\appendix

\section{Form equivalence}\label{app:notation}
In~\cite{angeris2020improved}, the trading function is defined as a function $\phi: \reals^n_+ \times \reals^n_+ \times \reals^n_+ \to 
\reals$, which maps the reserves, \emph{input trades}, and \emph{output trades} to a real value. In this paper, we do not make an explicit
distinction between the input and output trades; these are, instead, specified by the sign of the trade amounts $\Delta$ and $\Delta'$. 
In this case, we can make the following equivalences between the trade $(\Delta, \Delta')$ (the notation as used in this paper) and the input trade $\Delta^0 \in \reals_+^2$ and output trade $\Lambda^0 \in \reals_+^2$ as used in~\cite{angeris2020improved}, for $n=2$:
\[
\Delta^0 = ((-\Delta)_+, (\Delta')_+), \quad \Lambda^0 = ((\Delta)_+, (-\Delta')_+),
\]
while the reserves are simply $R^0 = (R, R')$. We then have
\[
\phi(R^0, \Delta^0, \Lambda^0) = \psi(R, R', \Delta, \Delta'),
\]
as expected. The update equations remain identical, since $R^0 \gets R^0 + \Delta - \Lambda$ is equivalent to $R \gets R - \Delta$
and $R' \gets R' + \Delta'$, which means that all results from~\cite{angeris2020improved} hold as stated.

\section{Lower bounds for portfolio value with fees}\label{app:pv-lower-bound}
The analysis of CFMMs with fees is often much harder than the analysis of fee-less CFMMs.
This construction gives a simple lower bound which shows that it often suffices to consider a fee-less CFMM with fees
taken separately; \ie, it often suffices to consider a CFMM where the fee is not reinvested into the reserves, but is
instead given to LPs directly.

\paragraph{Statement.} For simplicity, we will use the notation from~\cite{angeris2020improved}, which results in a simple proof
for any number of coins $n$. (See appendix~\ref{app:notation}.)

Let $\phi: \reals^n_+ \times \reals^n_+ \times \reals^n_+ \to \reals$ be a trading function for a CFMM that can be written as (with
some slight abuse of notation):
\[
\phi(R^0, \Delta^0, \Lambda^0) = \phi(R^0 + \gamma\Delta^0 - \Lambda^0),
\]
where $R^0 \in \reals_+^n$ are the reserves, $\Delta^0 \in \reals_+^n$ is the input trade, and $\Lambda^0 \in \reals_+^n$ is the output trade. We will assume that $\phi$ is increasing in its arguments, and let $0 < \gamma \le 1$ such that $(1-\gamma)$ is the fee taken for the CFMM. (See, \eg,~\cite[\S2.3.1]{angeris2020improved}.) In this case, we will consider the resulting portfolio value of an LP
at some cost vector $c \in \reals_+^n$. We will then show that the portfolio value of the LP after any feasible trade $(\Delta^0, \Lambda^0)$ is at least as large as the equivalent portfolio value at the previous reserves with an extra factor of $(1-\gamma)c^T\Delta$.

\paragraph{Proof.} The proof is nearly immediate. Let $R^1 = R^0 + \Delta^0 - \Lambda^0$ be the post-trade reserves and $R^0$ be the pre-trade reserves, then
\[
c^TR^1 = c^T(R^0 + \Delta^0 - \Lambda^0) = c^T(R^0 + \gamma \Delta^0 - \Lambda^0) + (1-\gamma)c^T\Delta^0 \ge c^TR^\star + (1-\gamma)c^T\Delta^0,
\]
where $R^\star$ is the solution to the fee-less portfolio-value problem~\cite[\S2.5]{angeris2020improved},
\begin{equation}\label{eq:pv}
p_{R^0}(c) = \inf_{\psi(R) \ge \psi(R^0)} c^TR,
\end{equation}
with variable $R \in \reals_+^n$. Note that the second inequality follows since, by definition, $(\Delta^0, \Lambda^0)$ is a feasible 
trade only when
\[
\psi(R^0 + \gamma\Delta^0 - \Lambda^0) \ge \psi(R^0),
\]
and so $R = R^0 + \gamma\Delta^0 - \Lambda^0$ is a feasible point for the portfolio-value problem~\eqref{eq:pv}.
Repeatedly applying this statement to any number of feasible trades
$(\Delta^k, \Lambda^k)$ yields the following lower bound for the portfolio value at time $k$:
\[
c^TR^k \ge p_{R^0}(c) + (1-\gamma)\sum_{k=1}^n c^T\Delta^k.
\]
In many cases, because we are interested in finding a lower bound to the portfolio value of liquidity providers,
it will often suffice to use this statement in order to achieve a reasonable lower bound. This allows us to side-step
the potentially very complicated analysis of CFMMs with fees and the fees' interactions with the CFMM's reserves.

\section{Relationship between $g$ and curvature of $\psi$}\label{app:curvcfmm}
From equation~\eqref{eq:g_def}, we can write,
\[
g(\Delta) = \frac{\frac{\partial \psi}{\partial\Delta}}{\frac{\partial \psi}{\partial\Delta'}} = \frac{\frac{\partial \psi}{\partial\Delta'} \frac{\partial\Delta'}{\partial \Delta}}{\frac{\partial \psi}{\partial\Delta'}} = \frac{d \Delta'}{d \Delta}
\]
Given a CFMM invariant function, given initial reserves, $\psi(\Delta, \Delta')$, we can write:
\[
\frac{\partial \psi}{\partial \Delta} = \frac{\partial \psi}{\partial \Delta'} \frac{\partial \Delta'}{\partial \Delta} = \frac{\partial \psi}{\partial \Delta'} g(\Delta)
\]
Therefore, $g(\Delta) = \left(\frac{\partial \psi}{\partial \Delta'}\right)^{-1} \frac{\partial \psi}{\partial \Delta}$.
Thus $\mu$-stability condition, for sufficiently smooth $g$, relies on the first derivative of $g$:
\begin{align*}
\frac{d g}{d \Delta} &= -\left(\frac{\partial \psi}{\partial \Delta'}\right)^{-2} \left(\frac{\partial \psi}{\partial \Delta \partial \Delta'}\right)\left(\frac{\partial \psi}{\partial \Delta}\right) + \left(\frac{\partial \psi}{\partial \Delta'}\right)^{-1} \frac{\partial^2 \psi}{\partial \Delta^2} \\ 
&= \left(\frac{\partial \psi}{\partial \Delta'}\right)^{-1} \left(g \frac{\partial\psi}{\partial\Delta\partial\Delta'} + \frac{\partial^2 \psi}{\partial \Delta^2}\right)
\end{align*}
From~\cite[Prop.\ 3.1]{goldman2005curvature}, we see that for an implicit function $F : \reals^2 \rightarrow \reals$, the Gaussian curvature $\kappa_F$ is defined as,
\begin{equation}\label{eq:gauss_planar}
\kappa_F = \frac{(\partial_2F)^2 (\partial_1^2F) - 2 (\partial_1F) (\partial_2F) (\partial_1\partial_2F) + (\partial_1F)^2 (\partial_2^2F)}{((\partial_1F)^2 + (\partial_2F)^2)^{3/2}}
\end{equation}
where $\partial_i F$ is the $i$th partial derivative of $F$.
Using implicit substitution (\eg, writing $\Delta'(\Delta))$ and substituting it into eq. \eqref{eq:gauss_planar}), we can see that the two formulas are equivalent.

\section{Curve's price impact function is convex}\label{app:curve_convex}
In this section, we claim that if the following assumption on the reserve sizes holds, then Curve's impact function is convex:
\begin{equation}\label{eq:reserve_constraint}
\rb - \db > \frac{1}{\ra}
\end{equation}
Recall that for Curve, the trading function is defined for $\alpha, \beta > 0$ as,
\[
\psi(\da, \db) = \alpha(\ra - \da + \rb + \db) + \frac{\beta}{(\ra-\da)(\rb + \db)} - k
\]

From~\cite[Prop. 3.8]{shifrin2015differential}, a curve defined via a sufficiently smooth implicit function $F(x, y) = 0$ is convex if and only if its Gaussian curvature (\ie, equation \eqref{eq:gauss_planar}) is non-negative.
Therefore, the claim of convexity is equivalent to showing that $\kappa_{\psi} \geq 0$.
Since the denominator of equation \eqref{eq:gauss_planar} is non-negative, we only need the numerator to be positive, \eg,:
\begin{equation}\label{eq:pos_condition}
(\partial_2\psi)^2 (\partial_1^2\psi^2) + (\partial_1\psi)^2 (\partial_2^2\psi)^2 \geq 2 (\partial_1\psi)(\partial_2\psi) (\partial_1\partial_2\psi)
\end{equation}
Using the definition of $\psi$, we have
\[
\begin{aligned}
    \partial_1\psi &= \alpha - \frac{\beta}{(R - \Delta)^2(R' + \Delta')} & \partial_2\psi &= -\alpha + \frac{\beta}{(R - \Delta)(R' + \Delta')^2} \\
    \partial_1^2\psi &= \frac{2\beta}{(R - \Delta)^3(R' + \Delta')} & \partial_2^2\psi &= \frac{2\beta}{(R - \Delta)(R' + \Delta')^3} \\
    \partial_1\partial_2\psi = \partial_2\partial_1\psi &= \frac{-\beta}{(R- \Delta)^2(R' + \Delta')^2} 
\end{aligned}
\]
Note that
\[
\begin{aligned}
    (R - \Delta)^5(R' + \Delta')^5(\partial_1\psi)^2 (\partial_2^2\psi) &= 2b(\alpha(R - \Delta)^2(R' + \Delta')-\beta)^2 \\
    (R - \Delta)^5(R' + \Delta')^5(\partial_2\psi)^2(\partial_1^2\psi) &= 2b(\alpha(R - \Delta)(R' + \Delta')^2-\beta)^2 \\
    (R - \Delta)^5(R' + \Delta')^5(\partial_1\psi) (\partial_2\psi) (\partial_1\partial_2\psi)&= \beta(\alpha(R - \Delta)^2(R' + \Delta')-\beta)(\alpha(R - \Delta)(R' + \Delta')^2-\beta)
\end{aligned}
\]
As $(R - \Delta)^5(R' + \Delta')^5 \geq 0$, combining results with~\eqref{eq:pos_condition} yields the positivity condition
\[
\begin{aligned}
2((\alpha(R - \Delta)^2(R' + \Delta')-\beta)^2 &+ (\alpha(R - \Delta)(R' + \Delta')^2-\beta)^2) \\
&\geq \\
(\alpha(R - \Delta)^2(R' + \Delta')-\beta)&(\alpha(R - \Delta)(R' + \Delta')^2-\beta)
\end{aligned}
\]
Let $A = \alpha(R - \Delta)^2(R' + \Delta')$ and $B = \alpha(R - \Delta)(R' + \Delta')^2$.
Then this condition can be rewritten as
\[
2((A-b)^2+(B-a)^2) \geq (A-b)(B-a)
\]
Dividing through by $(A-\beta)(B-\alpha)$ gives the final condition
\[
2\left(\frac{A-\beta}{B-\alpha} + \frac{B-\alpha}{A-\beta}\right) \geq 1
\]
Provided that $A > \beta$ and $B > \alpha$, then this condition is always true.
The first condition is equivalent to $(R - \Delta)^2(R' + \Delta') > \frac{\alpha}{\beta}$ and the second condition is equivalent to $(R - \Delta)(R' + \Delta')^2 > \frac{\beta}{\alpha}$.
Combining these gives the condition $(R - \Delta)(R' + \Delta') > 1$.
Minorizing $R - \Delta$ by $R$ is equivalent to the assumption \eqref{eq:reserve_constraint}, proving the claim.

\section{Portfolio Greeks for $n$ coin CFMM}\label{app:ncoinpv}
In the $n$ coin scenario, the portfolio value~\cite{angeris2020improved} under no arbitrage is:
\[
P_V = \sum_{i=1}^n \frac{\partial_i\Psi(R)}{\partial_n\Psi(R)} R_i = \sum_{i=1}^n m_i R_i
\]
where $R \in \reals_+^n$ is the reserve for each asset and we assume asset $n$ is the num\'eraire, such that $m_n = 1$.
Implicit differentiation of $\Psi$ gives,
\[
\frac{d\Psi(R)}{dm_i} = \sum_{i=1}^n \partial_i\Psi(R) \frac{dR_i}{dm_i} = 0
\]
Dividing through by $\partial_n\Psi(R)$ gives
\[
\frac{1}{\Psi_n(R)}\frac{d\Psi(R)}{dm_i} = \sum_{i=1}^n m_i \frac{dR_i}{dm_i} = 0
\]
This yields
\[
(P_\Delta)_i = \frac{dP_V}{dm_i} = R_i + m_i\frac{dR_i}{dm_i} + \sum_{j\neq i} \frac{dR_j}{dm_i} = R_i + \sum_{j\neq i} (m_j - 1) \frac{dR_j}{dm_i}
\]
and 
\[
(P_\Gamma)_{ii} = \frac{d^2P_V}{dm_i^2} = \frac{dR_i}{dm_i} + \sum_{j\neq i} (m_j-1) \frac{d^2 R_j}{dm_i^2}
\]
Note that when $m_j \approx 1$, which is what happens in stablecoin--stablecoin trading, $P_\Delta$ and $P_\Gamma$ resemble the two-asset trading hedges. 

\section{Conjecture: Delta hedging impermanent Loss}
The linear bounds of \eqref{eq:delta_hedge} suggest that one can hedge $P_{\Delta}$ if $\frac{d\da}{dm}$ and $\frac{d\db}{dm}$ can be replicated using options on impact. We note that the results presented here do not hold as stated since the assumption that the
market price is decreasing, used in the definitions of $\mu$ and $\kappa$, is broken here. We suspect that more general results
of this form do hold in practice, but leave this as an open question.

First, note that $\frac{d\da}{dm}$ is the inverse of the price impact function $\frac{dm}{d\da}$.
Suppose that an LP wants to hedge their impermanent loss when the realized price impact is greater than a fixed quantity $\xi$.
For instance, an LP in a stablecoin pool might believe that the price of the assets never deviate by more than 10\% and fixed $\xi = 0.1$.
Mathematically, $\xi$ corresponds to an lower bound on the impact, e.g. $\frac{dm}{d\da} \geq \xi$.
Recall that the Carr-Madan expansion \cite[App. 1]{carr2001towards} says that any payoff $f \in L^2(\reals_+)$ can be expressed, for $\kappa > 0$ as
\[
f(F) = f(\kappa) + f'(\kappa)\left[(F-\kappa)_+ - (\kappa-F)_+ \right] + \int_{0}^{\kappa} f''(K) (K-F)_+ dK + \int_{\kappa}^{\infty} f''(K) (F-K)_+ dK 
\]
Consider the function $f(F) = \frac{1}{F}\ones_{F > \xi+\epsilon} \in L^2(\reals_+)$ for some $\epsilon > 0$.
Using the Carr-Madan expansion at $F = \frac{dm}{d\da}$ and $\kappa = \xi$, we have:
\[
\left(\frac{dm}{d\da}\right)^{-1} = \int_{\xi+\epsilon}^{\infty} \frac{2}{K^3} \left(\frac{dm}{d\da}-K\right)_+ dK
\]
This equation states that we can replicate the exposure $\frac{d\da}{dm}\ones_{\frac{d\da}{dm} \geq \xi}$ by holding a portfolio of $\frac{2}{K^3}$ call options at strike $K$ for all $K \in [\xi, \infty]$.
Thus to hedge, we short this portfolio.
Using put-call parity, this is equivalent to holding a quantity of the asset $\alpha$ and selling put options at strike prices $K$ greater than our cutoff, weighted by $\frac{2}{K^3}$.
This effectively provides a way for an LP to insure themselves against impermanent losses up to price impacts of size $\xi$ by selling a portfolio of put options.

\section{Multiperiod informed trading}\label{app:multiperiod_it}
Suppose that we have a discrete time series of probabilities $\alpha_t \in [1/2, 1)$.
At time $t$, an informed trader has probability $\alpha_t$ of predicting whether the true price $m(t+1)$ at time $t+1$ is equal to the trader's predicted price $p_{t+1}$.
Associated to each $\alpha_t$ is a pair of quantities $\da(t), \db(t) \in \reals$ that represent the trade quantities needed to take the true price $m(t)$ to the predicted price $p_{t+1}$.
We can compute these quantities implicitly via the following differential equation implied by \eqref{eq:g_def}:
\[
\partial_1 \Psi(\ra(t) + \da, \rb(t) - \db) = p_{t+1} \partial_2 \Psi(\ra(t) + \da, \rb(t) - \db)
\]
To execute this in practice, an informed trader would need compute the optimal quantities $\da^*, \db^*$ that satisfy these questions.
However, given that $\Psi$ can be quite complicated, such a trader would likely have to rely on approximate methods.
In particular, it is likely that greedy methods like gradient descent would be employed given that there are latency constraints in realistic settings.

How would we compute these optimal quantities to trade?
To utilize a local method like gradient descent, we first need to compute an objective function $h(\da, \db)$ that we minimize using $\partial_{\da}h, \partial_{\db}h$.
In light of the above equation, is natural to define the following objective function
\[
h(\da, -\db, p) = \partial_1 \Psi(\ra(t) + \da, \rb(t) - \db) - p \partial_2 \Psi(\ra(t) + \da, \rb(t) - \db)
\]
Let us assume that the informed trader is computationally constrained and attempts to approximate $\da, \db$ via gradient descent:
\begin{align}
\da(t)_n &= \da(t)_{n-1} - \eta_{\alpha} \partial_1 h(\da, \db, p) \\
\db(t)_n &= \db(t)_{n-1} + \eta_{\beta} \partial_2 h(\da, \db, p)
\end{align}
This recursion is the \emph{gradient descent-ascent algorithm} (GDA)~\cite{lin2019gradient} or forward-backward iteration~\cite{chen1997convergence}.
Note that due to the natural constraint of increasing one quantity while reducing another, we take gradients in the opposite directions for each quantity.
Moreover, this is how a number of CFMMs compute trade quantities in practice (\eg, Curve, where it has been the source of a number of vulnerabilities~\cite{egorov_vuln_2020, zeitz_2020}).

One natural question to ask is: how many time steps $n$ does an informed trader need to run this algorithm to compute the optimal trade quantities up to some prescribed error?
It turns out that this is directly related to the curvature of the CFMM.
Suppose that this algorithm is run for a maximum of $T > 0$ time steps.
Then we can write a recurrence for the expected reserve quantities as a function of time:
\begin{align}\label{eq:bucket-recursion}
E[\ra(t+1) | \ra(t)] &= \ra(t) + \alpha_t \da(t)_T + (1-\alpha_t) \hat{\Delta}(t) \\ 
E[\rb(t+1) | \rb(t)] &= \rb(t) + \alpha_t \db(t)_T + (1-\alpha_t) \hat{\Delta}'(t) 
\end{align}
where the $\hat{\Delta}$ are the quantities that would need to trade if an oracle provided the optimal quantities to move from $m(t)$ to $m(t+1)$.
In particular, the quantities $\hat{\Delta}, \hat{\Delta}'$ are the no-arbitrage quantities that are traded when a roundtrip trade is made by an arbitrageur and the LP books a profit (akin to eq. \eqref{eq:trade-bound}).

If the informed trader has a very high amount of edge or information (e.g. $\liminf_{t\rightarrow\infty}\alpha_t = 1$), then \eqref{eq:bucket-recursion} is completely dominated by the GDA time steps of the informed trader.
From~\cite[Theorem 1]{lin2019gradient}, we find the result that when the GDA steps dominate the evolution of \eqref{eq:bucket-recursion}, then if $T = \Omega\left(\frac{\mu^2}{\kappa}\right)$, the expected reserves will be very close to the optimal reserves $\da^*, \db^*$, e.g. 
\[
\forall \epsilon > 0, \exists t'(\epsilon) \in O(T) \text{ such that }|\da(t)_{T+t'(\epsilon)} - \da^*| < \epsilon
\]
In particular, \cite{lin2019gradient} uses effectively the same definition of curvature for solving minimax problems using GDA.
We note that their results depend weakly on dimension, so that this recursion can be extended to informed traders interacting with $n$-asset CFMM markets. 

This illustrates that the curvature of a CFMM also controls the computational complexity that an informed trader needs to act on perfect information (\eg, $\alpha_t \approx 1)$ and trade with a CFMM.
This formulation was inspired by constructions in robust machine learning that resemble adversarial information aggregation in a CFMM.
We conjecture that this provides a way to extend some of our results to higher dimensions.

\section{Optimal Balancer yield farming}\label{app:optimal_bal_yf}
Suppose that we have two Balancer pools consisting of two assets that have the same spot price $p$ at time $t'$.
Pool $i$ for $i \in \{1, 2\}$ has reserves $\ra^i, \rb^i \in \reals_+$ and exponent $\tau_i \in (0,1)$ so that the trading function for the $i$th pool is
\[
\Psi_i(\ra^i, \rb^i) = (\ra^i)^{\tau_i}(\rb^i)^{1-\tau_i}
\]
By definition, the spot price equivalence means that:
\begin{equation} \label{eq:spot_initial}
p = \frac{\ra^1}{\rb^1}\left(\frac{1-\tau_1}{\tau_1}\right) = \frac{\ra^2}{\rb^2}\left(\frac{1-\tau_2}{\tau_2}\right)
\end{equation}
Let $p_i(t)$ be the price of the $i$th pool at time $t$ so that $p_1(t') = p_2(t') = p$.
Assume that a liquidity provider owns the same fraction $b \in [0, 1]$ of each pool and that at time $t'+1$ a trade of size $\Delta$ from $x$ to $y$ occurs.
In the feeless case, this gives a quantity change $\tilde{\Delta}_i$ to each pool, where $\tilde{\Delta}_i$ is implicitly specified via the CFMM constraints
\[
k_i = (\ra^i)^{\tau_i} (\rb^i)^{1-\tau_i} = (x_i-\Delta)^{\tau_i} (y_i + \tilde{\Delta}_i)^{1-\tau_i}
\]
Let $P_V(\ra^i, \rb^i, t)$ be the portfolio value of pool $i$ at time $t$ in units of $x$.
Then we have,
\begin{align}
    P_V(\ra^i, \rb^i, t) &= b \left(\ra^i + p_i(t') \rb^i \right) = b \ra^i \left(1 + \frac{1-\tau_i}{\tau_i}\right) = \frac{b}{\tau_i} \ra^i \\
    P_V(\ra^i, \rb^i, t'+1) &= \frac{b}{\tau_i} (\ra^i - \Delta)
\end{align}
This implies that if $\ra^1 = \ra^2$, then $\tau_1 > \tau_2$ implies $P_V(\ra^1, \rb^1, t'+1) < P_V(\ra^2, rb^2, t'+1)$.
Less formally, this says that losses to portfolio value, given an equal reserve of $x$ are higher for less sharply curved Balancer pool (\eg, $\tau_1$ loses more than $\tau_2$).
If we wanted to incentivize liquidity in pool $\tau_1$ by issuing new $x$ to liquidity providers, how much would we have to pay?
This value is exactly equal to $P_V(\ra^2, \rb^2, t'+1) - P_V(\ra^1, \rb^1, t'+1)$, which represents the `excess loss' covered by printing token $\alpha$ (we're assuming that's the portfolio num\'eraire).
Using this equation we have:
\begin{align*}
\delta(t'+1) = \frac{P_V(\ra^2, \rb^2, t'+1) - P_V(\ra^1, \rb^1, t'+1)}{b} &= \frac{1}{\tau_2}(\ra^2 - \Delta) - \frac{1}{\tau_1} (\ra^1 - \Delta) \\
&= \frac{\ra^2}{\tau_2} - \frac{\ra^1}{\tau_1} + \Delta\left(\frac{1}{\tau_1}-\frac{1}{\tau_2}\right) \\ 
&=\left(\frac{\rb^2}{\tau_2}\right)\left(1 - \left(\frac{\rb^1}{\rb^2}\right)
\frac{1-\tau_2}{1-\tau_1}\right) \\
&+ \Delta\left(\frac{1}{\tau_1}-\frac{1}{\tau_2}\right)
\end{align*}
Thus, if $\delta(t'+1)$ is paid out pro-rata to holders via a Synthetix-like yield farming mechanism, we can encourage users to stay in the worse pool.
This is an improved formula over those used by pro-active market makers, such as Dodo~\cite{dodo_curve}, which don't try to perform this accounting on a trade-by-trade basis and instead rely on an oracle.
You can effectively force the user to lock liquidity into pool 1 from height $t'$ to $t'+k$ and then pay out $\sum_{i=0}^k \delta(t'+i)$ upon redemption.

\section{Conjecture: yield farming is superhedging}\label{app:superreplication}
Before making the comparison to superhedging in discrete time~\cite[\S3.4]{carr2018convex}, we will demonstrate that equations \eqref{eq:il_subsidy} and \eqref{eq:il_growth} can be recast as an optimization problem.
Recall that for a function $f$, $g$ is a subgradient of $f$ at $x$, \eg, $g \in \partial f(x)$ if $\forall y \in \dom(f)$, $f(y) - f(x) \geq g^T(y-x)$.
Let $P_{\varphi}(R, c)$ be the portfolio value for an LP at price $p$, reserves $R$ for trading function $\varphi$.
Equation \ref{eq:il_subsidy} can be restated in terms of portfolio value as,
\[
P_{\varphi}(R, m_0^s) - P_{\varphi}(R, m_0^e) \geq -\frac{\mu}{\kappa}(m_0^s - m_0^e)
\]
Let $f(p) = P_{\varphi}(R, p)$.
Then this condition is equivalent to $-\frac{\mu}{\kappa} \in \partial f(m_0^e)$.
Similarly, equation \eqref{eq:il_growth} corresponds to showing any admissible subsidy $R_{\ell}$ must satisfy $0 \in R_{\ell} + \partial f(m_0^e)$.
This connects the subsidy to an optimization problem, as optima for a function $f$ are found by showing that $0\in\partial f$ \cite[\S1.3]{carr2018convex}.
These subgradient conditions illustrate that there is a connection between the yield farming subsidy and optimizing changes in portfolio value.
Since the change in portfolio value is equivalent to solving a dual optimization problem \cite{angeris2020improved}, this shows that yield farming subsidies are equivalent to bounding the payoff an LP engenders under an arbitrage trade by a simpler curvature dependent payoff.
In mathematical finance, bounding complex payoffs with simpler payoffs is known as \emph{superhedging} \cite[\S3.4]{carr2018convex}.

Suppose, instead, that we start with the opposite problem of trying to find a CFMM trading function $\Psi$ that is equivalent to given CFMM trading function $\phi$ and a subsidy $R_{\ell}$. 
Another way to arrive at equation \eqref{eq:il_subsidy} is to consider the set of CFMMs that can have a valid portfolio value at $(R + (R_{\ell},0), p)$ and find an upper bound for this portfolio value.
More formally, suppose that we define the following set of trading functions:
\[
\mathcal{P}(R, R_{\ell}, R', m', \mu, \kappa) = \{\varphi: (R + R_{\ell}, R', m') \in \dom(P_{\varphi}), \varphi \text{ is } \mu \text{-stable and } \kappa\text{-liquid}  \} 
\]
We can consider $\mathcal{P}(R, R_{\ell}, R', m', \mu, \kappa)$ to be a subset of the set of closed, proper, l.s.c. convex functions~\cite{angeris2020improved}.
If we can find a quasiconvex function $\psi$ such that
\begin{equation}\label{eq:superhedge}
\psi = \sup \mathcal{P}(R, R_{\ell}, R', m', \mu, \kappa)
\end{equation}
then $\psi$ must satisfy \eqref{eq:il_subsidy} as per the discussion in \S\ref{sec:yield_farming_returns}.
With some slight modifications, trading functions can be put in bijective correspondence with portfolio values, which are contingent claims.
Finding a contingent claim whose payoff is the supremum over a set of admissible contingent claims is known as a superhedge~\cite[\S3.4]{carr2018convex},~\cite{cheridito2017duality}.
Both~\cite{carr2018convex, cheridito2017duality} map superhedging to a convex dual problem analogous to finding portfolio value that is dual to \eqref{eq:superhedge}.
These formulations are very similar to the relationship between trading function and portfolio value from \cite{angeris2020improved}.
While traditional superhedging involves taking a supremum over a set of equivalent martingale measures, we are instead taking a supremum over a set of contingent claims whose curvatures are constrained.
There is some literature on superhedging over sets of contingent claim payoffs that are not equivalent martingale measures (\eg,~\cite{schal1999martingale}), however, none of the dual frameworks known to the authors map cleanly to our definitions of curvature.
We conjecture a curvature claim result analogous to~\cite[Prop. 2.3]{cheridito2017duality} exists.

\end{document}